\documentclass{WileyMSP-template}

\newcommand{\um}{\textmu{}m} 
\newcommand{\ul}{\textmu{}l} 

\usepackage{amsmath} 
\usepackage{xcolor} 
\usepackage[sort&compress,numbers]{natbib} 
\usepackage[labeled]{multibib}
\newcites{Main}{References}
\newcites{SI}{Supplementary References}

\begin{document}

\justifying
\setlength{\parindent}{0pt}

\title{Anomalous Dynamics of Superparamagnetic Colloidal Microrobots with Tailored Statistics}

\maketitle

\author{Alessia~Gentili,}
\author{Rainer~Klages,}
\author{Giorgio~Volpe*}

\dedication{}

\begin{affiliations}
Dr. Alessia Gentili\\
Department of Chemistry, University College London, London, WC1H 0AJ, United Kingdom.\\

Dr. Rainer~Klages\\
Centre for Complex Systems, School of Mathematical Sciences, Queen Mary University of London, London, E1 4NS, United Kingdom.\\
London Mathematical Laboratory, London, W6 8RH, United Kingdom.\\

Prof. Giorgio~Volpe\\
Department of Chemistry, University College London, London, WC1H 0AJ, United Kingdom.\\
Email Address: g.volpe@ucl.ac.uk

\end{affiliations}


\keywords{Microrobots, Active Colloids, Programmable Navigation, Stochastic Dynamics, Anomalous Diffusion}

\begin{abstract}

Living organisms have developed advanced motion strategies for efficient space exploration, serving as inspiration for the movements of microrobots. These real-life strategies often involve anomalous dynamics displaying random movement patterns that deviate from Brownian motion. Despite their biological inspiration, autonomous stochastic navigation strategies of current microrobots remain much less versatile than those of their living counterparts. Supported by theoretical reasoning, this work demonstrates superparamagnetic colloidal microrobots with fully customizable stochastic dynamics displaying the entire spectrum of anomalous diffusion, from subdiffusion to superdiffusion, across statistically significant spatial and temporal scales (covering at least two decades). By simultaneously tuning microrobots' step-length distribution and, critically, their velocity autocorrelation function with magnetic fields, fundamental anomalous dynamics are reproduced with tailored properties mimicking L\'evy walks and fractional Brownian motion. These findings pave the way for programmable microrobotic systems that replicate optimal stochastic navigation strategies found in nature for applications in medical robotics and environmental remediation.

\end{abstract}


\section{Introduction}
Living organisms have evolved efficient locomotion strategies to navigate complex landscapes, search their surroundings, and improve their fitness \citeMain{VLRS11,GMO25}. Selecting an optimal navigation strategy maximizes their ability to locate resources, reach targets, and evade threats \citeMain{VLRS11,GMO25}. Often, optimal strategies yield deviations from normal diffusion known as anomalous diffusion \citeMain{KRS08}. These processes are characterized by a non-linear power-law scaling of the mean squared displacement in time, ${\rm MSD}(t) \sim t^\mu$, where $\mu$ is the anomalous diffusion exponent, including superdiffusion ($\mu > 1$) and subdiffusion ($\mu < 1$), as opposed to normal diffusion ($\mu = 1$) \citeMain{MeKl00}. Popular stochastic models describing anomalous dynamics in random navigation problems are (superdiffusive) Lévy walks, featuring heavy-tailed step-length distributions \citeMain{ZDK15,volpe2017topography}, and fractional Brownian motion, showing long-range correlations with both superdiffusion and subdiffusion 
\citeMain{mandelbrot1968fractional,KKK21}.

Inspired by these biological scenarios \citeMain{palagi2018bioinspired}, self-propelled nano- and microrobots have been designed for targeted applications in, e.g., nanomedicine \citeMain{ruiz2025micro} and environmental remediation \citeMain{urso2023smart}.
Among these engineered systems, active colloids are widely recognized as synthetic models for living matter \citeMain{bishop2023active,volpe2024roadmap}, with significant potential for microrobotic applications due to their simplicity, versatility and ease of fabrication \citeMain{liu2023colloidal}. 

Although advanced autonomous stochastic navigation strategies displaying anomalous dynamics have been successfully implemented and validated in macroscale robotics \citeMain{fujisawa2013levy,dimidov2016random}, hardware miniaturization constraints have posed significant hurdles to implement the same on smaller scales. Beyond numerical studies \citeMain{golestanian2009anomalous,shaebani2014anomalous, volpe2017topography,goswami2024anomalous,sevilla2024anomalous,nasiri2024smart}, enhanced diffusion ($\mu = 1$) and 
directed motion ($\mu \to 2$) continue to be the dominant types of fully autonomous navigation mechanisms for active colloids \citeMain{bechinger2016active,Lowen2020,gao2022micro}. Attempts at more advanced navigation strategies require information to be stored in the environment \citeMain{nakayama2023tunable, dias2023environmental} or external feedback loops to correct and steer trajectories in real time \citeMain{mano2017optimal,bauerle2018self,yigit2019programmable,tal2020experimental,sprenger2020active,alvarez2021reconfigurable, goyal2025externally}, for example to implement reinforcement learning-based approaches \citeMain{muinos2021reinforcement, khadka2018active, heuthe2024counterfactual, loffler2023collective}. However, autonomous stochastic strategies resembling anomalous diffusion patterns as in living organisms remain elusive for active colloids, where, only on rare occasions, short trajectories compatible with Lévy walks have been observed with limited statistics and without robust control over long-term dynamics or precise control of the anomalous diffusion exponent \citeMain{KPV19, akella2020levy}.

Here, unlike previous experiments, we demonstrate superparamagnetic colloidal microrobots driven by external magnetic fields that move according to fundamental anomalous diffusion patterns with fully tailored statistics spanning the entire spectrum of anomalous diffusion, from subdiffusion ($\mu < 1$) to superdiffusion ($\mu > 1$), and over statistically significant temporal and spatial scales (covering at least two decades). Supported by theoretical reasoning, we achieve fine control over the microrobots' long-term dynamics by simultaneously tuning their step-length distribution and, critically, their velocity autocorrelation function. Thanks to this fine control, our microrobots describe two-dimensional trajectories displaying anomalous dynamics compatible with Lévy walks and fractional Brownian motion with tailored anomalous diffusion exponents, hence better mimicking natural stochastic navigation patterns \citeMain{VLRS11,GMO25}.  

\section{Results}

\subsection{Anomalous dynamics of colloidal microrobots in the comoving frame}

\begin{figure}[h!]
\centering
\includegraphics[width = \textwidth]{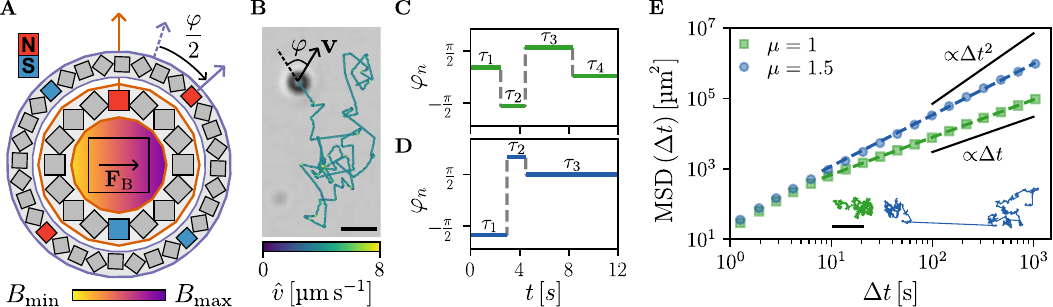}
\caption{\textbf{Anomalous diffusion of superparamagnetic colloidal microrobots in the comoving frame.} (\textbf{A}) Two concentric Halbach cylinders (inner dipole; outer quadrupole) generate a linear magnetic field $\mathbf{B}$ (color gradient) and a constant force $\mathbf{F_\mathrm{B}}$ (black arrow) on a sample of colloidal microrobots (black rectangle). The cylinders’ magnetic north (N, red) and south (S, blue) are shown. Rotating the quadrupole (purple arrows) around the fixed dipole (orange arrow) reorients $\mathbf{F}_\mathrm{B}$ and the microrobots' direction. Two consecutive rotations define their turning angle $\varphi$ as twice the quadrupole rotation angle.
(\textbf{B}) Trajectory of a colloidal microrobot moving at approximately constant speed ($4.6~\pm0.8$~\um{}~$\rm s^{-1}$) in a comoving frame defined by its velocity vector $\mathbf{v}$ and $\varphi$. Scale bar: 25~\um.
(\textbf{C}-\textbf{D}) Bespoke sequences $\varphi_n$ in time $t$ of uniformly distributed $\varphi$ in $[-\pi, \pi)$, obtained by sampling the quadrupole rotation time $\tau_n$ (solid lines) from (\textbf{C}) a half-Gaussian and (\textbf{D}) a power-law (anomalous exponent $\mu = 1.5$) distribution.
(\textbf{E}) Time-averaged mean squared displacements (${\rm MSD}$, symbols) and  trajectories (inset) of microrobots yielding long-time normal diffusion ($\mu = 1$) and superdiffusion ($\mu = 1.5$) as confirmed by a logarithmic curve fit (dashed lines). Diffusive ($\propto \Delta t$) and ballistic ($\propto \Delta t^2$) slopes shown for reference. Scale bar: 1 mm.}
\label{fig:Fig1}
\end{figure}

In Fig. \ref{fig:Fig1}, we show typical trajectories of  microrobots yielding anomalous dynamics. Our microrobots are colloidal superparamagnetic silica spheres of diameter $13.8~\pm~0.4$~\um{} driven by external planar rotating magnetic fields (Experimental Section). We use two concentric Halbach cylinders \citeMain{blumler2023practical}, a dipole and a quadrupole, to generate a constant magnetic field gradient $|\nabla|\mathbf{B}||\sim 0.9~\, {\rm T \,~m^{-1}}$ 
(Fig. \ref{fig:Fig1}A and Fig. \ref{fig:Fig_Halbach}, Table \ref{tab_magnets}, Experimental Section). This gradient translates into a constant magnetic force $\mathbf{F}_\mathrm{B} = |\mathbf{m}|\boldsymbol{\nabla}|\mathbf{B}|$, where $\mathbf{m}$ is the particles' magnetic moment, that drives the colloidal microrobots at constant speed in its direction, $v_{\rm c} = |\mathbf{v}| =  \frac{\mathbf{F}_\mathrm{B}}{6\pi\eta R}$ (Experimental Section), where $\mathbf{v}$ is the particle's velocity, $R$ its radius, and $\eta$ the fluid's viscosity \citeMain{baun2017permanent}. By rotating the quadrupole around the fixed dipole at discrete times $t_n$ with $n$ an integer (Fig. \ref{fig:Fig1}A, Experimental Section), we can reorient $\mathbf{F}_\mathrm{B}$ and, hence, the microrobots' motion direction to generate extended tailored trajectories in a two-dimensional comoving frame (Fig. \ref{fig:Fig1}B). This is a coordinate frame that defines the microrobot's motion in terms of its speed $|\mathbf{v}|\ge0$ and turning angle $\varphi \in[-\pi,\pi)$, where here only the latter changes at times $t_n$ (Figs. \ref{fig:Fig1}C-D, Supplementary Text). Originally introduced by Ross and Pearson about a century ago (Supplementary Text), this coordinate frame has been used to analyze and model foraging organisms in movement ecology \citeMain{CPB08} and to formulate advanced stochastic processes such as two-dimensional L\'evy walks \citeMain{zaburdaev2016superdiffusive,volpe2017topography}. Arguably, the comoving frame is also the most natural one to study the (anomalous) dynamics of active agents, such as living organisms and robots, driven by an internal source of randomness generated by the agents themselves \citeMain{lenz2013constructing,dimidov2016random}. 
Generating stochastic dynamics in a comoving frame implies a fundamental change of perspective compared to defining stochastic processes in a more standard fixed Cartesian frame (Supplementary Text). Assuming overdamped dynamics, two decoupled time-discrete stochastic equations allow us to formulate our microrobot's two-dimensional motion in the comoving frame as (Supplementary Text) \citeMain{lenz2013constructing}
\begin{eqnarray} 
  	 \varphi_n &=& \xi_{\varphi,n} \label{eq:overd1d}\\
  	 v_n &=& \xi_{v,n}\label{eq:overd2d}
\end{eqnarray}
where $\xi_{\varphi,n}$ and $\xi_{v,n}$ represent  arbitrarily complex noise terms driving each coordinate's dynamics sampled at (non-necessarily equally spaced) discrete times $t_n = t_{n-1}$ + $\tau_n$ with $\tau_n$ the quadrupole rotation time (Experimental Section, Fig. \ref{fig:Fig1}). Given constant speed, the microrobot runs a distance $\ell_n = v_{\rm c} \tau_n$ ballistically during flight time $\tau_n$. If we sample $\tau_n$ from an arbitrary noise distribution $\xi_{\tau,n}$, we can replace Eq.~\ref{eq:overd2d} with (Supplementary Text)
\begin{eqnarray} 
  	 \ell_n &=& v_{\rm c}\xi_{\tau,n} \label{eq:overd2d2}
\end{eqnarray}
where the step length and flight time distributions are related by $\xi_{\ell,n} = v_{\rm c}\xi_{\tau,n}$.
By defining appropriate distributions for $\xi_{\varphi,n}$ (through the quadrupole rotation angle) and $\xi_{\tau,n}$ (through the quadrupole rotation time) and by sampling independent and identically distributed random variables from these distributions in time, random walks with different statistical properties can be generated in the comoving frame experimentally under the constraint of constant speed (Fig. \ref{fig:Fig1}, Experimental Section, Supplementary Text). For example, sampling $\xi_{\varphi,n}$ from the uniform distribution on the circle, our microrobots can describe trajectories yielding long-time normal diffusion or superdiffusion (Fig. \ref{fig:Fig1}E) when $\xi_{\tau,n}$ is sampled from either a half-Gaussian distribution \citeMain{Sant20} (Fig. \ref{fig:Fig1}C) or a power law distribution with exponent $\alpha = 3 - \mu$ as in L\'evy walks \citeMain{zaburdaev2016superdiffusive} (Fig. \ref{fig:Fig1}D for $\alpha = \mu = 1.5$) (Experimental Section, Supplementary Text). 
Unlike the normal diffusion case, the superdiffusive trajectory shows that the microrobot performs occasional long jumps displaying the spatial features typical of L\'evy walks. The random nature of these long jumps is confirmed by the mean displacement of the microrobot in time being nearly zero ($< 0.07 R$, with $R$ the particle's radius). The narrow distributions of the instantaneous speed $\hat{v}$ for each trajectory confirm that our microrobots move at an approximately constant speed (Fig. \ref{fig:Fig_Instant_Speed}), 
validating the use of Eqs.~\ref{eq:overd1d} and \ref{eq:overd2d2} to formulate their motion. Since the microrobot's reorientation timescale is controlled by the implemented stochastic dynamics rather than rotational Brownian motion, our colloidal microrobots have a controllable variable average step length (from 12 \um{} for normal diffusion to 16 \um{} for superdiffusion) even at constant speed, 
unlike systems governed by enhanced diffusion \citeMain{bechinger2016active}. A long-time fit of the time-averaged mean squared displacement (MSD) calculated from each trajectory (for $\Delta t > 8$ s, i.e. above the short-time persistence of the trajectory due to the magnetic drive) \citeMain{volpe2014simulation} confirms the two desired diffusion regimes over two decades (Fig. \ref{fig:Fig1}E, Experimental Section). From the fits, we indeed estimate the anomalous diffusion exponents to be $\hat\mu = 1.0653 \pm 0.0002$ and $\hat\mu = 1.5473 \pm 0.0004$, respectively. Our microrobots therefore travel distances two orders of magnitude longer than their own size while reliably maintaining the desired anomalous dynamics (Experimental Section).

\begin{figure}[h!]
\centering
\includegraphics[width = \textwidth]{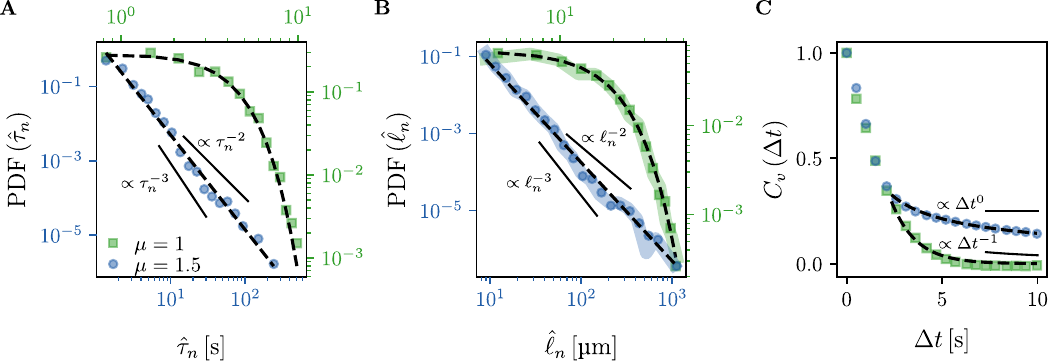}
\caption{{\bf Tailored diffusion statistics of superparamagnetic colloidal microrobots}. (\textbf{A}-\textbf{B}) Probability distribution functions (PDF)  of (\textbf{A}) flight times $\hat{\tau}_n$ between turns and (\textbf{B}) step lengths $\hat{\ell}_n$ extracted from our microrobots' trajectories for the cases of normal diffusion ($\mu = 1$, green squares) and superdiffusion ($\mu = 1.5$, blue circles) as in Fig. \ref{fig:Fig1}E.
(\textbf{C}) Respective normalized time-averaged velocity autocorrelation $C_v$ as a function of lag time $\Delta t$. In (\textbf{A}-\textbf{C}), fitting curves to the functions (dashed lines) show exponential and power-law scaling respectively consistent with normal diffusion ($\mu = 1$) and superdiffusion ($\mu = 1.5$, Table \ref{tab_Stats_Fig3}). The thick background lines in (\textbf{B}) represent ${\rm PDF} (\langle \hat{v} \rangle\hat\tau_{n})$, showing that ${\rm PDF} (\hat\ell_n) \sim {\rm PDF} (\langle \hat{v} \rangle\hat\tau_{n})$ with $\langle \hat{v} \rangle$ the microrobot's measured mean instantaneous speed (Fig. \ref{fig:Fig_Instant_Speed}). 
The axis colors in (\textbf{A}-\textbf{B}) reflect those of the respective distributions. Diffusive (\textbf{A}: $\propto \tau^{-3}_n$; \textbf{B}: $\propto \ell_n^{-3}$; \textbf{C}: $\propto \Delta t^{-2}$) and ballistic (\textbf{A}: $\propto \tau^{-2}_n$; \textbf{B}: $\propto \ell^{-2}_n$; \textbf{C}: $\propto \Delta t^{0}$) limits shown for reference.}
\label{fig:Fig2}
\end{figure}

\subsection{Analysis of microrobots' trajectory statistics}

A deeper analysis of the experimental trajectories' statistics, based on their segmentation with the detected turning points (Fig. \ref{fig:Fig_Turning_Points}, Experimental Section), further confirms that the final microrobots' dynamics are consistent with the desired diffusion regimes (Fig. \ref{fig:Fig2}).
Beyond the mean squared displacements (Fig. \ref{fig:Fig1}E), we can extract the flight times $\hat{\tau}_n$, the step lengths $\hat\ell_n$ and the turning angles $\hat{\varphi}_n$ of our microrobots directly from each trajectory (Experimental Section). The step lengths $\hat{\ell}_n$ depend linearly on the respective flight times $\hat{\tau}_n$ (Fig. \ref{fig:Fig_Time_Step_Speed}), thus providing an independent confirmation of the microrobots' approximately constant speed. 
The probability distribution functions (PDFs) of $\hat{\varphi}_n$, $\hat{\tau}_n$ and $\hat\ell_n$  (Fig. \ref{fig:Fig2}A-B and Fig. \ref{fig:Fig_Angle_Time_Distributions}) confirm that the microrobots are reproducing the desired distributions as defined by the quadrupole's rotation angle and time (Experimental Section): for both trajectories in Fig. \ref{fig:Fig1}E, the distribution of the turning angle is uniform on the circle (Fig. \ref{fig:Fig_Angle_Time_Distributions}A-B), thus matching the intended sampling specified by the quadrupole's rotation in line with Eq. \ref{eq:overd1d} (i.e., ${\rm PDF}(\hat\varphi_n) \sim {\rm PDF}(\varphi_n)$, Supplementary Text); both distributions of $\hat{\ell}_n$ and $\hat{\tau}_n$ show exponential and power-law scaling (with $\hat\mu \approx 1.5$ over two decades, Table \ref{tab_Stats_Fig3}), respectively consistent with normal diffusion ($\mu = 1$) and superdiffusion for $\mu = 1.5$ (Figs. \ref{fig:Fig2}A-B and Fig. \ref{fig:Fig_Angle_Time_Distributions}C) \citeMain{zaburdaev2016superdiffusive}, thus matching the intended sampling specified by the quadrupole’s rotation times in line with Eq.~\ref{eq:overd2d2} (i.e., ${ \rm PDF}(\hat{\ell}_n) \sim  {\rm PDF}(\langle \hat{v} \rangle \hat{\tau}_n) \sim {\rm PDF}(\langle \hat{v} \rangle \tau_{n})$, Supplementary Text).
The time-averaged experimental velocity auto-correlation function $C_v(\Delta t)$ at lag times $\Delta t$ is also in agreement with theoretical expectations for the two regimes (Fig. \ref{fig:Fig2}C, Experimental Section) \citeMain{BoGe90,Lowen2020}. The tail of this function decays as an exponential for the diffusive case due to the short-term persistence in the magnetic field \citeMain{Lowen2020} and, asymptotically, as a power-law ($\sim \Delta t^{\hat\mu -2}$ with $\hat\mu = 1.434 \pm 0.005$, Table \ref{tab_Stats_Fig3}) for the superdiffusive case, as expected for unbiased L\'evy walks \citeMain{ZDK15}.

\begin{figure}[ht]
\centering
\includegraphics[width = \textwidth]{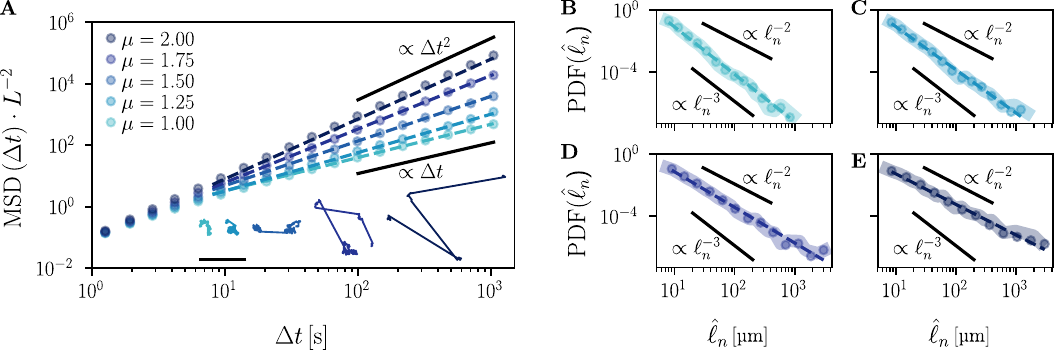}
\caption{{\bf Tailoring microrobots' superdiffusion by controlling step-length distributions}. (\textbf{A}) Normalized time-averaged mean squared displacements (${\rm MSD}$, dots) yielding long-time superdiffusion for microrobots' trajectories (inset) generated according to Eqs.~\ref{eq:overd1d} and \ref{eq:overd2d2} by sampling the turning angle $\varphi_n$ from the uniform distribution on the circle and the flight time $\tau_n$ from power-law distributions of varying exponent $\alpha = 3 - \mu$ between the normal diffusive ($\mu = 1$) and ballistic ($\mu = 2$) limits. Case for $\mu = 1.5$ as in Fig. \ref{fig:Fig1}E. Fit lines (dashed lines) confirm the different superdiffusive regimes (Table \ref{tab_Stats_Fig3}). The MSDs are normalized to the square of each microrobot's short-term drift distance $L$ in the driving magnetic field. Scale bar: 5 mm.
(\textbf{B-E}) Probability distribution functions (PDF, dots) of experimental step lengths $\hat{\ell}_n$ for (\textbf{B}) $\mu = 1$, (\textbf{C}) $\mu = 1.25$, (\textbf{D}) $\mu = 1.75$ and (\textbf{E}) $\mu = 2$. ${\rm PDF} (\hat\ell_n) \sim {\rm PDF} (\langle \hat{v} \rangle\,\hat\tau_n)$ (thick background lines). Case for $\mu = 1.5$ in Fig. \ref{fig:Fig2}B. 
Fit lines (dashed lines) show power-law scalings ($\sim \hat{\ell}_n^{\hat{\mu}-4}$) consistent with the desired ground-truth values of $\mu$ (Table \ref{tab_Stats_Fig3}). Diffusive (\textbf{A}: $\propto \Delta t$; \textbf{B-E}: $\propto \ell_n^{-3}$) and ballistic (\textbf{A}: $\propto \Delta t^2$; \textbf{B-E}: $\propto \ell^{-2}_n$) limits shown for reference.} \label{fig:Fig3}
\end{figure}

\subsection{Tuning microrobots' step-length distributions}
The statistics in Figs. \ref{fig:Fig1} and \ref{fig:Fig2} demonstrate that our colloidal microrobots can perform superdiffusion consistent with a L\'evy walk model over two decades in space and time. For our approach to be truly versatile, control over the anomalous diffusion exponent $\mu$ is desirable, since this parameter allows us to tune the average step length of our microrobots even at constant speed. 
Fig. \ref{fig:Fig3} shows the possibility of tuning the values of $\mu$ between the diffusive ($\mu = 1$) and ballistic ($\mu = 2$) limits by controlling the distributions of the step lengths $\ell_n$. By sampling $\varphi_n$ from the uniform distribution on the circle and $\tau_{n}$ from power-law distributions of varying exponent $\alpha = 3 - \mu$ (Fig. \ref{fig:Fig_Angle_Time_Distributions_Levy}) \citeMain{zaburdaev2016superdiffusive}, our microrobots can describe trajectories in the comoving frame according to Eqs.~\ref{eq:overd1d} and \ref{eq:overd2d2} yielding different regimes of superdiffusion in a controllable way under the experimental constraint of constant speed (Table \ref{tab_Stats_Fig3}). Fig. \ref{fig:Fig3}A (inset) shows example trajectories for different values of $\mu$: as the anomalous diffusion exponent increases, the microrobots tend to move ballistically over longer distances before a random change in orientation occurs. As expected for unbiased L\'evy walks \citeMain{ZDK15}, the mean value of the displacements for each trajectory is near zero ($< 0.07\,R$). Importantly, the four independent measurements $\hat{\mu}$ of the anomalous diffusion exponent obtained from fitting mean squared displacements (Fig. \ref{fig:Fig3}A, $\sim \Delta t^{\hat{\mu}}$), probability distribution functions of flight times $\hat\tau_n$ (Fig. \ref{fig:Fig_Angle_Time_Distributions_Levy}, $\sim \hat{\tau}_n^{\hat{\mu}-4}$), probability distribution functions of step lengths $\hat\ell_n$ (Fig. \ref{fig:Fig3}B-E, $\sim \hat{\ell}_n^{\hat{\mu}-4}$) and velocity autocorrelation functions (Fig. \ref{fig:Cv_Levy}, $\sim \Delta t^{\hat{\mu}-2}$) all scale in agreement with theoretical expectations for L\'evy walks at the respective ground-truth value of $\mu$ (Table \ref{tab_Stats_Fig3}) \citeMain{ZDK15}. Consistent with this scaling, the mean step length of the microrobots also increases with increasing $\mu$, from $\approx 11$ \um{} at $\mu = 1$ to $\approx 118$ \um{} at $\mu = 2$.

The ability to precisely tune the anomalous diffusion exponent is particularly beneficial, as different environments may require distinct optimal search strategies. For example, Lévy walks have been shown to be highly efficient in search problems \citeMain{VLRS11}, but their optimality can depend on the specific value of $\mu$ according to environmental characteristics \citeMain{volpe2017topography, viswanathan2008levy}.

\begin{figure}[ht]
\centering
\includegraphics[width = \textwidth]{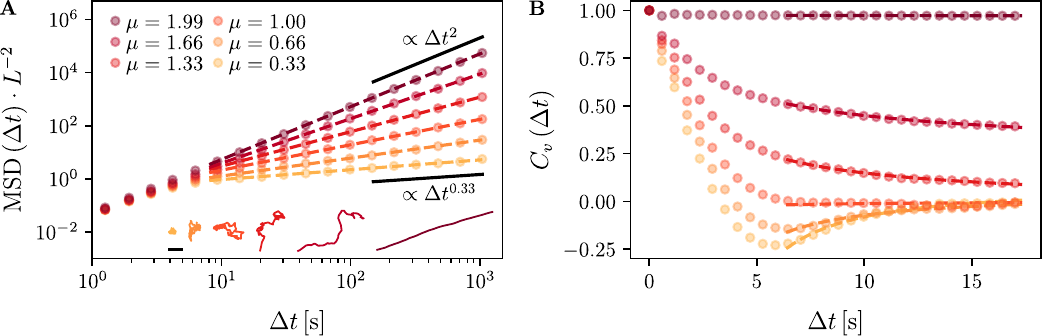}
\caption{{\bf Tailoring anomalous diffusion by controlling the microrobots' velocity autocorrelation function}. (\textbf{A}) Normalized time-averaged mean squared displacements (${\rm MSD}$, dots) yielding different anomalous dynamics at long times, from subdiffusion ($\mu < 1$) to superdiffusion ($\mu > 1$) through normal diffusion ($\mu = 1$), for microrobots' trajectories (inset) consistent with fractional Brownian motion (Experimental Section, Supplementary Text). Fit lines (dashed lines) confirm the different anomalous diffusion regimes (Table \ref{tab_Stats_Fig4}). The MSDs are normalized to the square of each microrobot's short-term drift distance $L$ in the driving magnetic field. Diffusive ($\propto \Delta t$) and ballistic ($\propto \Delta t^2$) limits shown for reference. Scale bar: 100~\um.
(\textbf{B}) Respective normalized time-averaged velocity autocorrelation functions $C_v$ (dots) as a function of lag time $\Delta t$ calculated from the trajectories in (\textbf{A}) for different ground-truth values of $\mu$. Fitting the tail of the data with a quadratic polynomial scaled by a power law (dashed lines) confirms the asymptotic scaling characteristic of fractional Brownian motion at different values of $\mu$ ($\sim \hat{\mu}(\hat{\mu}-1)\Delta t^{\hat{\mu}-2}$, Table \ref{tab_Stats_Fig4}).} \label{fig:Fig4}
\end{figure}

\subsection{Tuning microrobots' velocity autocorrelation functions}
The superdiffusive dynamics discussed so far belong to a class of memoryless anomalous dynamics, which means that the microrobot's persistent motion does not depend on its past steps in the trajectory. Introducing memory into diffusive dynamics enables the adoption of both persistent and antipersistent motions, the trade-off of which can optimize time efficiency versus area coverage in space exploration tasks \citeMain{GMO25}. Such anomalous diffusion dynamics arise when the agent's displacements are not independent but correlated in time. 
A famous example for this type of dynamics is fractional Brownian motion where the driving noise is no longer white but colored \citeMain{mandelbrot1968fractional}. This stochastic process can generate the whole spectrum of anomalous diffusion under parameter variation, from subdiffusion ($\mu < 1$) to superdiffusion ($\mu >1$) through normal diffusion ($\mu = 1$) \citeMain{mandelbrot1968fractional}. Experimentally, we implemented microrobots whose motion satisfies Eqs.~\ref{eq:overd1d} and \ref{eq:overd2d2} by pregenerating sequences of flight times $\tau_n$ and turning angles $\varphi_n$ that yield an analogue of two-dimensional fractional Brownian motion in the comoving frame under the constraint of constant speed (Experimental Section, Supplementary Text). In analogy to the transformation between L\'evy flights and walks, we refer to this constant-speed process as \textit{fractional Brownian walks}. Fig. \ref{fig:Fig4} shows that, as $\mu$ increases, the corresponding trajectories (Fig. \ref{fig:Fig4}A, inset) become less localized and more ballistic. The microrobots' tendency to turn backward (negative persistence) reduces in favor of its forward propagation (positive persistence) (polar plots, Fig. \ref{fig:Polar_plots_fbm}). A long-time fit of the time-averaged mean squared displacements (MSDs) calculated from each trajectory (Fig. \ref{fig:Fig4}A, Table \ref{tab_Stats_Fig4}) confirms the shift from subdiffusion (sublinear MSD, $\mu < 1$) to superdiffusion (superlinear MSD, $\mu > 1$) through normal diffusion (linear MSD, $\mu = 1$), when the microrobots' velocities show a transition from negative to positive correlations (Fig. \ref{fig:Fig4}B, Table \ref{tab_Stats_Fig4}) in agreement with theoretical expectations (Experimental Section, Supplementary Text) \citeMain{jeon2010fractional}. Therefore, these two consistent independent measurements (i.e., by fitting the MSD and $C_v$) of the anomalous diffusion exponent associated with each trajectory strongly support the tailored generation of different types of anomalous diffusion dynamics, compatible with fractional Brownian motion, for our colloidal superparamagnetic microrobots by the spatio-temporal control of their turning angles and flight times in the comoving frame (Table \ref{tab_Stats_Fig4}).

\section{Conclusion}

We have demonstrated colloidal superparamagnetic microrobots capable of programmable anomalous dynamics compatible with two-dimensional models of normal diffusion \citeMain{Lowen2020}, L\'evy walks \citeMain{ZDK15}, and fractional Brownian motion \citeMain{mandelbrot1968fractional}. Supported by theoretical reasoning, we have implemented these anomalous dynamics in a comoving frame, i.e. a frame moving and rotating with the microrobot, directly capturing the motion from the perspective of the active agent itself. We validated these dynamics over statistically relevant temporal and spatial scales by precisely tuning two key experimental parameters (i.e., the microrobot's turning angles and flight times).
Our approach enables the computationally efficient motion planning of colloidal microrobots capable of autonomous navigation based on diverse advanced random strategies without requiring onboard circuitry \citeMain{palagi2018bioinspired} or the implementation of any external feedback based on the position and velocity of the microrobots \citeMain{mano2017optimal,bauerle2018self,yigit2019programmable,sprenger2020active,alvarez2021reconfigurable, goyal2025externally}. Such autonomous random navigation strategies can prove beneficial in exploring complex unknown environments, where deterministic strategies may struggle and a stochastic approach may be preferred \citeMain{savoie2019robot, GMO25}.
By implementing motion from the microrobot's perspective, magnetic actuation in the comoving frame ensures biocompatibility, external programmability, and remote control, making it particularly suitable for manipulation and navigation tasks in therapeutic applications \citeMain{yang2020magnetic} and environmental remediation \citeMain{li2022self}. Beyond advancing the capabilities for autonomous navigation of microrobots, we also anticipate that our framework will provide a robust experimental platform for validating theoretical and predictive models of anomalous diffusion dynamics in active matter, thus contributing to deepen our general understanding of anomalous diffusion processes across various fields and scales, from the life sciences to macroscopic natural and human processes \citeMain{waigh2023heterogeneous,Vilk22,munoz2021objective}.


\section{Experimental Section}

\threesubsection{Materials}\\
Glass microscopy slides (25~mm x 75~mm x 1~mm, Epredia) and glass coverslips (24~mm x 24~mm x 0.14~mm) for sample preparation were purchased from Thermo Fisher and VWR, respectively. The following chemicals were purchased and used as received: acetone ($\geq 99.8~\%$, Sigma-Aldrich), ethanol ($\geq 99.8~\%$, Fisher Scientific), ethylene glycol (Sigma-Aldrich), Tween 20 (Sigma-Aldrich). Deionised (DI) water ($\geq 18\;\rm~M\Omega\;~cm$, type II Water) was collected from a Milli-Q purification system. Aqueous colloidal dispersions (5\% w/v) of superparamagnetic silica (SiO$_2$) particles were purchased from Microparticles GmbH. Parafilm (Bemis Parafilm M Laboratory Wrapping Film), used as spacer for the sample chamber, was purchased from Fisher Scientific. Two-part epoxy glue (Gorilla Epoxy) for sealing the samples was purchased from RS Components. The neodymium magnets used to build the Halbach cylinders were purchased from K\&J Magnetics, Inc. (B666-N52) and supermagnete (W-07-N). PA 2200 (nylon powder) was used to 3D-print the encasing of the magnets for the cylinders.

\threesubsection{Colloidal dispersion}\\
As our microrobots, we used superparamagnetic silica (SiO$_2$) colloidal particles with a diameter of $13.8 \pm 0.4$ ~\um, an iron oxide content greater than 5 wt.~\%, and a high density of approximately 1.5~$\mathrm{g}\,~\mathrm{cm}^{-3}$, as estimated by the manufacturer. Before each experiment, we gradually diluted the original batch dispersion in a 50~\% ethylene glycol and 50\% DI water solution by volume to achieve very low particle concentrations ($<$~10$^{-5}$~w/v\%) and avoid interparticle interactions in a magnetic field. The viscosity of this solution is approximately four times that of pure DI water \citeMain{sun2003density} to reduce the particles' speed when exposed to the high magnetic fields generated by the Halbach cylinders (Fig. \ref{fig:Fig1}). Typical P\'eclet numbers range between 4000 and 6000. Such high values indicate that Brownian motion is negligible and directed motion dominates the short-term dynamics of our particles \citeMain{bechinger2016active}.
To prevent the particles from sticking to the glass slides during experiments, we added small traces ($<$ 0.002 v/v\%) of a 10\% Tween 20 aqueous solution to the final dispersion. By preventing sticking and by increasing viscosity to reduce occurrences of particles exiting the field of view, we were able to control their dynamics for up to 9 hours.

\threesubsection{Sample chamber}\\
We confined 62~\ul{} of the colloidal dispersion within a quasi-two-dimensional chamber assembled from a microscope slide (bottom layer, cut to approximately 25~mm x 28~mm x 1~mm) and a coverslip (top layer) using two strips of melted parafilm as spacers to obtain a thickness of $\approx$ 20~\um. First, we cleaned both the glass slide and coverslip by sequentially immersing them in Coplin jars containing acetone, ethanol and DI water in an ultrasonic bath for 5, 10 and 15 minutes, respectively. Blowing the slide and coverslip dry with nitrogen gas removed excess water. We then placed two strips of parafilm (approximately 25~mm x 5~mm each) at opposite edges of the glass slide and let them melt on a hotplate at 60$^{\circ}$C, near the melting point of parafilm. Once the parafilm turned transparent (after about 3 minutes), we placed the coverslip on top, and applied slight pressure using the flat tip of a pair of tweezers to close the chamber. After cooling and loading the particles' dispersion, we sealed the chamber with two-part epoxy glue, and let it cure and rest for at least 20 minutes before each experiment.

\threesubsection{Magnetic fields with Halbach cylinders}\\
We generated the constant magnetic field gradient $\nabla |\mathbf{B}|$ needed to drive our superparamagnetic colloidal microrobots at constant speed $v_{\rm c} = |\mathbf{v}|$ in the sample plane using two concentrical Halbach cylinders (Fig. \ref{fig:Fig1}A and Fig. \ref{fig:Fig_Halbach}): an inner dipole (Fig. \ref{fig:Fig_Halbach}A) surrounded by an outer quadrupole (Fig. \ref{fig:Fig_Halbach}B) \citeMain{blumler2023practical}. These cylinders, constituted by circular arrays of permanent magnets (Fig. \ref{fig:Fig_Halbach}C), can produce controlled magnetic fields entirely within their core while canceling it on the outside. In our case, the axis of the cylinders is aligned along the direction ($z$ in Fig. \ref{fig:Fig_Halbach}) perpendicular to the sample plane ($xy$ in Fig. \ref{fig:Fig_Halbach}). The inner Halbach cylinder produces a strong homogeneous dipolar magnetic field $\mathbf{B}^{\rm D}$ in this plane with constant intensity $B_\mathrm{0}$ along the $y$-axis (Fig. \ref{fig:Fig_Halbach}A, D-F), maximizing the magnetic moment $\mathbf{m}$ of the superparamagnetic particles and aligning it along the field lines. The outer cylinder generates a weaker quadrupolar magnetic field $\mathbf{B}^{\rm Q}$ consisting of two orthogonal linear components in space (i.e., each with a constant derivative of magnitude $G$) (Fig. \ref{fig:Fig_Halbach}B,G-I). When the two arrays are coaxially aligned, the resulting field $\textbf{B}(\textbf{r})$ at position $\textbf{r} = (x,y)$ is linear in space and given by \citeMain{blumler2023practical} 
\begin{equation}\label{eq:B_rotation} 
\mathbf{B}(\mathbf{r}) = \mathbf{B}^{\rm D}(\mathbf{r}) + \mathbf{B}^{\rm Q}(\mathbf{r}) = B_0 
\begin{bmatrix} 
0 \\ 1 
\end{bmatrix} + G \begin{bmatrix} -\cos2\beta & \sin2\beta \\ \sin2\beta & \cos2\beta
\end{bmatrix} 
\begin{bmatrix} 
x \\ y 
\end{bmatrix}, 
\end{equation}
where $2\beta$ is the angle of rotation of the magnetic field gradient induced by a $\beta$ rotation of the quadrupole around the dipole. The direction of the gradient can therefore be adjusted by rotating the quadrupole with respect to the dipole (Fig. \ref{fig:Fig1}A), and the difference $\Delta \beta$ between two consecutive rotations of the quadrupole defines the microrobot's turning angle $\varphi$ as $\varphi = 2\Delta \beta$ (Fig. \ref{fig:Fig1}A-B). As a result of combining a strong homogeneous dipolar magnetic field with a weaker constantly graded one, our microrobots move in a well-defined, spatially independent and adjustable direction defined only by the component of the gradient parallel to $\mathbf{B}^{\rm D}$ (the $y$-component in Eq. \ref{eq:B_rotation}) \citeMain{baun2017permanent}.

We implemented a discrete version of the Halbach dipole with radius $r_{\rm c} = 30.05 \, {\rm mm}$ using $k = 16$ cubic neodymium magnets (grade N52, remanence $B_{\rm R} \sim 1.48 \, {\rm T}$, relative permeability $\mu_{\rm R} = 1.05$, side length $a_{\rm m} = 9.5 \, {\rm mm}$) with $B_0$ given by \citeMain{blumler2023practical}
\begin{align}\label{eq:Bd}
B_0 = B_\mathrm{R}\ln\left(\frac{r_{\mathrm{out}}}{r_{\mathrm{in}}}\right)\frac{1}{\sqrt{\mu_\mathrm{R}}}\frac{\sin(2\pi/k)}{2\pi/k}\frac{k\,a_{\mathrm{m}}}{\pi(r^2_{\mathrm{out}} - r^2_{\mathrm{in}})}\frac{h(6r_{\mathrm{c}}^2+h^2)}{(4r_{\mathrm{c}}^2+h^2)^{3/2}} \sim 85\,\rm mT
\end{align}
where $r_\mathrm{in} = 23.3 \, {\rm mm}$ and $r_\mathrm{out} = 36.8 \, {\rm mm}$ are the cylinder's inner and outer radii, respectively (Fig. \ref{fig:Fig_Halbach}C), and $h = a_{\rm m}$ its height. The cylinder radius $r_{\rm c}$ can be calculated from these two values as their average $r_\mathrm{c} = (r_\mathrm{in}+r_\mathrm{out})/2$ and was chosen to be more than double the entire sample's size to prevent edge effects due to field inhomogeneities nearer to the magnets. Moreover, we implemented the Halbach dipolar cylinder with a vertical stack of two identical circular arrays separated by 15.3~mm \citeMain{soltner2010dipolar}, to reduce field inhomogeneity in the $z$-direction, thus minimizing any possible vertical magnetic drift of the colloids. Stacking the two arrays also produces an expected increase in field intensity by a factor of 1.351 with respect to the prediction in Eq. \ref{eq:Bd}.  We confirmed this increase with a Gaussmeter (Lake Shore Cryotronics, Inc., Model 420), measuring an average magnetic field of $111.38\pm0.66$~mT (Fig. \ref{fig:Fig_Halbach}D-F). 

Similarly, we implemented a discrete version of the Halbach quadrupole with a larger radius $r_{\rm c} = 46.7 \,~{\rm mm}$ using $k = 32$ cubic neodymium magnets (grade N42, remanence $B_{\rm R} \sim 1.32\,~{\rm T}$, relative permeability $\mu_{\rm R} = 1.05$, side length $a_{\rm m} = 7\,~{\rm mm}$) with $G$ given by \citeMain{blumler2023practical}
\begin{align}\label{eq:Gq}
G = \frac{2B_{\mathrm{R}}}{\sqrt{\mu_{\mathrm{R}}}}\left(\frac{1}{r_{\mathrm{in}}} - \frac{1}{r_{\mathrm{out}}}\right)\frac{\sin(3\pi/k)}{3\pi/k}\frac{k\,a^2_{\mathrm{m}}}{\pi(r^2_{\mathrm{out}} - r^2_{\mathrm{in}})}\frac{h(h^4+10h^2r_{\mathrm{c}}^2+30r_{\mathrm{c}}^4)}{(4r_{\mathrm{c}}^2+h^2)^{5/2}} \sim0.9\,\mathrm{Tm^{-1}}.
\end{align}
This value was confirmed calculating the gradient ($0.98 \pm 0.08\,~\mathrm{Tm^{-1}}$) from the magnetic field intensities measured with the Gaussmeter (Fig. \ref{fig:Fig_Halbach}G-I). 

Table \ref{tab_magnets} summarizes all parameters used to implement both Halbach cylinders. All magnets were held in place side by side by plastic molds (one for the dipole and one for the quadrupole), which we 3D-printed using Selective Laser Sintering (SLS) technology.

\threesubsection{Experimental setup}\\
We recorded microrobots' trajectories using a custom-built inverted microscope. The sample rested in the region of homogeneous dipolar field at the center of the Halbach dipole, which was supported by four metal pillars (Thorlabs). The sample holder was uncoupled from the dipole support to reduce transmission of vibrations due to the rotation of the quadrupole around the dipole. For the same reason, the quadrupole was mounted on a third separate support, vertically centered at the sample level, and connected to a high-speed motorized rotational stage (Zaber, X-RSB060AD). To reorient the magnetic field gradient instantaneously with respect to the microrobots' dynamics, we rotated the quadrupole at constant angular speed ($60 \, {\rm rad \, s^{-1}}$). We chose this value because, during the time required by the quadrupole to complete the largest rotation in our experiments ($\Delta \beta = \pm\pi/2$), a microrobot traveled a distance comparable with our localization error on the determination of its centroid ($\approx$ 0.13~\um), thus with negligible influence on the final trajectory. To reduce vibrations, we ramped the quadrupole rotation up to (down from) its maximum speed with a constant angular acceleration (deceleration) of $60 \, {\rm rad \, s^{-2}}$.
For sample illumination, a monochromatic LED ($\lambda =$ 660 nm, Thorlabs, M660L4) equipped with an adjustable collimation adapter (Thorlabs, SM2F32-A) was mounted on this last support. To acquire long trajectories, we also uncoupled the imaging system from the part of the setup containing the sample. The imaging system was formed by two lenses projecting the image of the sample with a 4x magnification on a monochrome complementary metal–oxide–semiconductor (CMOS) camera (Thorlabs, DCC1545M). This system was mounted on a computer-controlled two-axis motorized translation stage (Thorlabs, PT1/M-Z8) to allow us to recenter the microrobot in the field of view of the camera (1.3 mm x 1.6 mm), thus avoiding that the particle exited it in long linear stretches of its motion. Videos of microrobots were recorded with a frame rate of $11.94$ frames per second (the inverse of the sampling time $\delta t$) using a custom MATLAB program that triggered the camera acquisition. The same program controlled the sequence of rotations of the quadrupole to implement bespoke patterns of anomalous diffusion and the translation of the imaging system based on the microrobot's position. During relatively long quadrupole rotation times ($>$ 15 s), video recording was temporarily interrupted to automatically recenter the microrobot to the field of view every 2 seconds before recording was resumed.
We translated the stage with a constant speed of $1\,\rm{mm}\,s^{-1}$, approximately 200 times faster than typical particles' speeds, i.e. almost instantaneously compared to the microrobots' dynamics (Fig. \ref{fig:Fig_Instant_Speed}). To reduce vibrations, we ramped the stage translation up to (down from) its maximum speed with a constant acceleration (deceleration) of $1\,\rm{mm}\,s^{-2}$. 
Full microrobots' trajectories were reconstructed by stitching together individual trajectories (Fig. \ref{fig:Fig_Traj_StitchingA}, see \emph{Trajectory stitching}) obtained from sequences of videos corresponding to each experiment using homemade Python scripts based on the \textsc{Trackpy} package \citeMain{allan_2023_7670439}. Like this, we were able to acquire trajectories over centimeter-long scales over extended periods of time (up to 9 hours).

\threesubsection{Trajectory stitching}\\
We reconstructed full microrobots' trajectories by stitching together individual trajectories from a sequence of $N$ consecutive videos. To facilitate stitching, any two consecutive videos respectively finished and started with an at least 1-s long portion of the same step length $\ell_n$ in the trajectory (Fig. \ref{fig:Fig_Traj_StitchingA}). These portions were reconnected by translating all the $i$ points $\mathbf{r}_i^{j+1}$ of the trajectory (with $j$ and $i$ both integers) defined in the coordinate system of the $(j+1)^{\rm th}$ video back to the reference system of the $j^{\rm th}$ video (Fig. \ref{fig:Fig_Traj_StitchingA}). The origin of the coordinate system associated to each video is at the center of its field of view. As the microrobot is moving ballistically at the time of its recentering, we also accounted for the additional distance it travels between recordings (Fig. \ref{fig:Fig_Traj_StitchingA}). The translation between the reference systems of two consecutive videos is then given by
\begin{eqnarray}\label{eq:stiching}
    \mathbf{r}_i^{j} = \mathbf{r}_i^{j+1} + \mathbf{r}^{j}_M \approx \big( \mathbf{r}_i^{j+1} - \mathbf{r}^{j+1}_0 \big) + \tau_{\delta} \langle \hat{v}_\ell \rangle \mathbf{u}_v + \mathbf{r}_M^{j}
\end{eqnarray}
where $\mathbf{r}^{j}_i$ and $\mathbf{r}_M^{j}$ are respectively the positions of $\mathbf{r}_i^{j+1}$ and of the last recorded point of the $j^{\rm th}$ video in its reference system, and $\mathbf{r}^{j+1}_0$ identifies the first particle's position of the $(j+1)^{\rm th}$ video in its reference system. If the microrobot is moving ballistically at approximately constant speed (as in our experiments, Fig. \ref{fig:Fig_Instant_Speed}), the vector $\mathbf{r}^{j+1}_0 \approx \tau_{\delta} \langle \hat{v}_\ell \rangle \mathbf{u}_v$, where $\tau_{\delta}$ is the time elapsed between recordings, $\langle \hat{v}_\ell \rangle$ is the average particle's speed in the two recorded portions of the step length being reconstructed, and $\mathbf{u}_v = \big({\rm cos} (\theta), {\rm sin}(\theta) \big)$ is the unitary vector in the current direction of motion, i.e. also defined by the same step length of the full trajectory which is being reconstructed. Finally, we linearly interpolated the displacement between the last point of the $j^{\rm th}$ video and the first point of the $({j+1})^{\rm th}$ video by resampling with our experimental sampling time $\delta t$ (the inverse of the frame rate). We repeated this procedure iteratively until the trajectory was fully reconstructed in the coordinate system of the first acquisition video. Fig. \ref{fig:Fig_Traj_StitchingB} verifies that any two reconnected portions in the reconstructed trajectory maintain the same direction of motion (Fig. \ref{fig:Fig_Traj_StitchingB}A) and the validity of the relationship $\mathbf{r}^{j+1}_0 \approx \tau_{\delta} \langle \hat{v}_\ell \rangle \mathbf{u}_v$ based on direct measurements of $\mathbf{r}^{j+1}_0$ from image analysis and measurements of stage displacements (Fig. \ref{fig:Fig_Traj_StitchingB}B), thus further confirming our constant speed approximation.

\threesubsection{Distributions of quadrupole rotation times and turning angles}\\
For trajectories yielding normal diffusion (Figs. \ref{fig:Fig1} and \ref{fig:Fig2}, $\mu = 1$), we numerically generated sequences of $N$ quadrupole rotation times $\tau_n$ drawn from a half-Gaussian distribution (Fig. \ref{fig:Fig_Angle_Time_Distributions}). The probability density function (PDF) of this distribution is 
\begin{equation}
\mathrm{PDF}(\tau) = \frac{e^{-\tau^2/16}}{2\sqrt{\pi}},\;\; \tau \geq 0
\end{equation}
For trajectories yielding L\'evy walks of exponent $\alpha = 3 - \mu$ (Figs. \ref{fig:Fig1}, \ref{fig:Fig2} and \ref{fig:Fig3}), sequences of $N$ quadrupole rotation times $\tau_n$ were numerically generated using the inverse method (figs. \ref{fig:Fig_Angle_Time_Distributions} and \ref{fig:Fig_Angle_Time_Distributions_Levy}) \citeMain{newman2005power}: by drawing $r$ as a random number from a uniform distribution in $[0,1)$, the variable $\tau = \tau_{\rm min}(1-r)^{-1/\alpha}$ follows a power-law distribution with exponent ($\alpha + 1$) and lower bound $\tau_{\rm min}$
\begin{equation}
\mathrm{PDF}(\tau) = C\tau^{-(\alpha + 1)},\;\; \tau \geq \tau_{\rm min} \label{eq:powertrunc}
\end{equation}
where $C = \alpha\tau^\alpha_{\rm min}\,$ is a normalization constant. A power-law distribution with a lower bound was preferred over an $\alpha$-stable L\'evy distribution to optimize experimental time by focusing directly on the tail of the distributions. For this purpose, we set $\tau_{\rm min} = 1 \, {\rm s}$ to facilitate the detection of turning points in the trajectories (see \emph{Turning point detection}).

Finally, we drew sequences of $N$ turning angles $\varphi_n$ from a uniform distribution over the half-open interval [$-\pi$, $\pi$) for both normal diffusion and L\'evy walks (figs. \ref{fig:Fig_Angle_Time_Distributions} and \ref{fig:Fig_Angle_Time_Distributions_Levy}). In all cases, we set $N$ so that the cumulative sum of all $\tau_n$ was at least 3-hours long to observe anomalous diffusion in experiments over at least two decades in space and time.  

For fractional Brownian motion, we generated sequences of $(\tau_n,\varphi_n)$ yielding a constant-speed analogue of this process in the comoving frame, which satisfies Eqs.~\ref{eq:overd1d} and \ref{eq:overd2d2}, by adopting the protocol detailed in the Supplementary Text. In analogy to the transformation between L\'evy flights and walk, we refer to realizations of this process as \emph{fractional Brownian walks}. Briefly, we transformed trajectories generated in simulations with a constant flight time $\tilde{\tau}_{\rm c}$ and non-constant Gaussian-distributed velocities $\tilde{v}_n$ in a Cartesian frame into trajectories of the same path topology with constant speed $v_{\rm c}$ and non-constant flight times $\tau_n$ in the comoving frame by doing the following (Supplementary Text): we first generated two-dimensional (time-discrete) Cartesian fractional Brownian motion from the Davies-Harte method \citeMain{davies1987tests} implemented in the Python package \textsc{stochastic} for all values of the anomalous exponent $\mu$ used in the experiments; in order to scale the average speed in simulations ($\tilde{v}_{\rm c}=\langle \tilde{v}_n \rangle = \sqrt{\frac{\pi}{2}}\sigma_{\tilde{v}}$ for $\sigma_{\tilde{v}} = 1$ \um{} $\rm{s}^{-1}$ and $\tilde{\tau}_{\rm c} = 1 \, \rm{s}$) to a representative a-priori estimate for the microrobot's experimental average speed $v_{\rm c} =4.5$ \um $\rm{s}^{-1}$, we then used  Eq.~\ref{eq:kappav} with a scale factor of $\kappa = 3.6$ (defined by Eq.~\ref{eq:lscale}), thus effectively matching our experimental length scales; finally, we transformed the scaled sequences of speeds and turning angles $(v_n,\varphi_n)$ associated to each trajectory into the corresponding $(\tau_n,\varphi_n)$ sequences for the rotation of the quadrupole implementing the transformation given by Eq. \ref{eq:vtscale}.

\threesubsection{MSD calculation and fitting}\\
For each trajectory, we calculated the time-averaged mean squared displacement (MSD) at discrete time lags $\Delta t = m \delta t$ (with $\delta t$ the experimental sampling time and $m$ an integer) as \citeMain{volpe2014simulation}
\begin{equation}
    \mathrm{MSD}(\Delta t) = \frac{1}{T-\Delta t} \sum_{t=\delta t}^{T-\Delta t} \left[ \big(x(t+\Delta t) - x(t)\big)^2 + \big(y(t+\Delta t) - y(t)\big)^2 \right]
\end{equation}
where $\mathbf{r} = (x,y)$ are the trajectory's coordinates sampled at time steps $t = p\delta t$ (with $p$ an integer), and $T = P\delta t$ (with $P = 12500$) is the total number of data points in the $\mathrm{MSD}$ calculations. We chose $T$ to be shorter than the trajectory length, but large enough to extract anomalous diffusion exponents by fitting the MSD over at least two decades with strong statistical reliability. The scaling exponent $\hat{\mu}$ of the $\mathrm{MSD}$ was estimated with a linear fit in log-log scale in the asymptotic limit (i.e., for $\Delta t > 8$ s, after the short-time persistence transition point, Figs. \ref{fig:Fig1}C, \ref{fig:Fig3}A and \ref{fig:Fig4}A). The reported uncertainty associated with the estimated exponent $\hat{\mu}$ (tables \ref{tab_Stats_Fig3} and \ref{tab_Stats_Fig4}) corresponds to one standard deviation of the fit parameter.

\threesubsection{Detection of turning points}\\
We identified turning points along microrobots' trajectories based on the detection of local extrema in their velocity. Given that our microrobots move at nearly constant speed (Fig. \ref{fig:Fig_Instant_Speed}), significant variations of this quantity should primarily reflect directional changes. Experimentally, we used variations of the absolute value of the acceleration magnitude gradient ($|\nabla|\textbf{a}||$, Fig. \ref{fig:Fig_Turning_Points}A) as a noise-robust empirical proxy to identify these directional changes. To further minimize the impact of the experimental noise, this time series was preprocessed with a Savitzky-Golay filter with a 5-point kernel \citeMain{orfanidis1995introduction}, implemented with the Python \textsc{scipy.signal.savgol\_filter} function. Prominent peaks were then identified using the Python \textsc{scipy.signal.find\_peaks} function \citeMain{virtanen2020scipy}.
We validated this method for the independent detection of the turning points directly from the acquired trajectories by comparing their predicted values $\hat{\tau}_n$ against the ground truth from the sequences $\tau_n$ of quadrupole rotations (Fig. \ref{fig:Fig_Turning_Points}B). For all trajectories, we achieved a $\mathrm{F_1}$ score of at least 0.83. Here, we computed the micro average of the $\mathrm{F_1}$ score using the Python \textsc{sklearn.metrics.f1\_score} function with a tolerance of five data points ($\sim 0.42$ s) \citeMain{pedregosa2011scikit}, i.e. we considered a predicted turning point a true positive if it was within five points of a ground truth value. 

\threesubsection{Experimental distributions of flight times and step lengths}\\
After identifying the turning points along each trajectory, we calculated the probability density functions (PDFs) of the flight times $\hat{\tau}_n$ and step lengths $\hat{\ell}_n$ of the particles (Figs. \ref{fig:Fig2}, \ref{fig:Fig3}, figs. \ref{fig:Fig_Angle_Time_Distributions}, \ref{fig:Fig_Angle_Time_Distributions_Levy} and \ref{fig:Fig_Time_Distributions_fBm}). For normal diffusion (Fig. \ref{fig:Fig2}A-B), the PDFs should decay exponentially in the long-time limit, which we verified by fitting them to a half-Gaussian function \citeMain{KRS08,ZDK15}. For trajectories yielding L\'evy walks (Figs. \ref{fig:Fig2}A-B and \ref{fig:Fig3}B-E), we verified their asymptotic power-law scaling and corresponding anomalous exponent $\mu$  with a linear fit of the distribution tails on log-log scale (Table \ref{tab_Stats_Fig3}) \citeMain{KRS08,ZDK15}. To increase tail statistics, we combined data from three different trajectories for each value of $\mu$. For fractional Brownian walks, the PDFs should follow a Rayleigh distribution independent of the anomalous diffusion exponent $\mu$ (Supplementary Text), as confirmed experimentally (Fig. \ref{fig:Fig_Time_Distributions_fBm}). 

\threesubsection{Experimental velocity autocorrelation functions}\\
For each trajectory, we calculated the normalized velocity autocorrelation function (VACF) as
\begin{equation}
C_v(\Delta t) = \frac{\langle \mathbf{v}(t)\cdot \mathbf{v}(t + \Delta t) \rangle}{\langle \mathbf{v}^2(t) \rangle}
\end{equation}
where $\mathbf{v}(t)$ is the instantaneous particle's velocity at time $t$, $\Delta t$ is the time lag for the calculation of the VACF and $\langle\dots\rangle$ indicates a time average. 
For normal diffusion, $C_v(\Delta t)$ follows an exponential decay as expected (Fig. \ref{fig:Fig2}C) \citeMain{ZDK15,Lowen2020}.   
For trajectories yielding L\'evy walks, we confirmed that $C_v(\Delta t)$ decays as a power law of consistent anomalous exponent $\mu$ asymptotically (Fig. \ref{fig:Cv_Levy}, Table \ref{tab_Stats_Fig3}) \citeMain{ZDK15}.
For fractional Brownian walks, we confirmed that $C_v(\Delta t)$ decays as the asymptotic functional form characteristic of this process given by $C_v(\Delta t) \simeq \frac{1}{2}\mu(\mu-1)\Delta t^{\mu-2}$ for each value of $\mu$ (Fig. \ref{fig:Fig4}B, Table \ref{tab_Stats_Fig4}) \citeMain{benelli2021sub}.

\newpage
\textbf{Supporting Information} \par 
Supporting Information is available from the Wiley Online Library or from the author.

\medskip
\textbf{Acknowledgements} \par 
A.G and G.V. acknowledge sponsorship for this work by the US Office of Naval Research Global (award no. N62909-18-1-2170).

\medskip
\textbf{Conflict of Interest} \par 
The authors declare no conflict of interest.

\medskip
\textbf{Author contributions} \par 
Author contributions are defined based on the CRediT (Contributor Roles Taxonomy) and listed alphabetically. Conceptualization: G.V. Data curation: A.G. Formal analysis: A.G., R.K., G.V. Funding acquisition: G.V. Investigation: A.G. Methodology: A.G., R.K., G.V. Project administration: G.V. Resources: G.V. Software: A.G., G.V. Supervision: G.V. Validation: A.G., G.V. Visualization: A.G. Writing -- original draft: G.V. Writing -- review and editing: All.

\medskip
\textbf{Data Availability Statement} \par 
All data supporting the findings of this study are available in the manuscript and its Supporting Information. Further data can be obtained from the corresponding author on reasonable request. Code in support of the finding of this study can be obtained from the corresponding author on reasonable request.

\medskip

%
\begin{raggedright}

\end{raggedright}




\clearpage


\renewcommand{\thefigure}{S\arabic{figure}}
\renewcommand{\thetable}{S\arabic{table}}
\renewcommand{\theequation}{S\arabic{equation}}
\renewcommand{\thepage}{S\arabic{page}}
\setcounter{figure}{0}
\setcounter{table}{0}
\setcounter{equation}{0}
\setcounter{page}{1} 

\renewcommand{\refname}{} 
\markboth{}{} 

\begin{center}
    {\LARGE\bfseries Supporting Information}\\[1.5em]

    {\large\bfseries Anomalous Dynamics of Superparamagnetic Colloidal Microrobots with Tailored Statistics}\\[1em]

    Alessia Gentili, Rainer Klages, Giorgio Volpe*\\[0.5em]
    \textit{*Corresponding author: g.volpe@ucl.ac.uk}
\end{center}

\section*{Supplementary Text: Stochastic processes in the comoving frame} \label{si:dcm}
Here, we provide the theoretical underpinning for our experiments. We first introduce the comoving frame and illustrate its use by a simple example. We then define equations for overdamped stochastic dynamics in it by implementing our experimental constraint of constant speed. Next, we lay out a protocol to define trajectories with constant speed in this reference frame compatible with fractional Brownian motion, which we refer to as \emph{fractional Brownian walks}. Finally, we outline the specific implementation of this protocol for our experiments.

\subsection*{The comoving frame} \label{si:cm}
Stochastic processes are typically defined in a fixed Cartesian
frame, as this can simplify their theoretical analysis. In contrast, our
experiments are better described by stochastic dynamics formulated in a comoving frame in two dimensions (Fig. \ref{fig:Fig1}). This is a coordinate frame attached to the center of mass of a moving particle, whose $x$-axis is aligned with the velocity vector $\mathbf{v}_{n-1}$ at any time step $t_{n-1}$, and the associated $y$-axis is perpendicular to it \citeSI{RBELS12}. It is thus corotating with the change of direction of the moving particle, and, hence, includes both translational and rotational movements. In this Cartesian comoving frame the velocity $\mathbf{v}_n$ at the next time step $t_n$ can be expressed in terms of its abscissa $v_{x,n}$ along the $x$-axis and its ordinate $v_{y,n}$ along the $y$-axis. However, it is more convenient to formulate the velocity $\mathbf{v}_n$ at the next time step $t_n$ in the comoving frame 
by its polar coordinates, speed $v_n$ and turning angle $\varphi_n$ (Fig. \ref{fig:Fig1}B). The pair $(\varphi_n,v_n)$ at time steps $t_n$ yields the coordinates of a particle in the comoving frame, and the corresponding time series fully determines its dynamics. 

Because of our experimental constant speed constraint, we can directly consider overdamped dynamics in the form of correlated random walks \citeMain{CPB08}. These are defined by the time-discrete equations of motion in the comoving frame (Eqs. \ref{eq:overd1d} and \ref{eq:overd2d} in the main text):
\begin{eqnarray} 
  	 \varphi_n &=& \xi_{\varphi,n} \label{eq:overd1ds} \\
  	 v_n &=& \xi_{v,n} \label{eq:overd2ds} \: .
\end{eqnarray}
The terms $\xi_{\varphi,n}$ and $\xi_{v,n}$ represent two noises driving our microrobots' experimental dynamics in this coordinate frame. In principle, these noises can be arbitrarily complex: they can depend on discrete time $t_n$; they can also be coupled by featuring prefactors that may depend on both state variables; furthermore, they may be correlated in time.

\subsection*{Random walks} \label{si:rw}
To illustrate this general framework, we can consider the two-dimensional time-discrete random walk put forward by Ross and Pearson about a century ago  (Fig. \ref{fig:Fig1}, $\mu = 1$) \citeMain{Ross04,Pea05}$,\,$\!\citeSI{Pea06}. If defined in a Cartesian coordinate frame, we can sample the two velocities, $v_x$ and $v_y$, along $x$ and $y$ as independently and identically distributed random variables from two corresponding probability distributions, $\xi_{v_x,n}\sim\rho(v_x)$ and $\xi_{v_y,n}\sim\rho(v_y)$, where here both distributions are symmetric Gaussians. Transforming these two velocities into polar coordinates of velocity orientation $\theta$ and speed $v$ by using conservation of probability \citeSI{Gard09},
\begin{equation}
    \rho(v_x)\rho(v_y)d v_x d v_y=\rho(\theta)\rho(v) d \theta d v\: , \label{eq:cop}
\end{equation}
yields a uniform distribution on the circle, $\rho(\theta)=(2\pi)^{-1}$, for the corresponding probability density of the orientation and a Rayleigh (two-dimensional Maxwell-Boltzmann) distribution for the corresponding probability density of the speed, 
\begin{equation}
    \rho(v)=\frac{v}{\sigma_v^2}e^{-\frac{v^2}{2\sigma_v^2}}\:. \label{eq:rayleigh}
\end{equation}
where $\sigma_v$ is the scale parameter of the distribution.
After further wrapping on the circle and due to Markovianity, $\rho(\theta)$ delivers an equally uniform turning angle distribution $\rho(\varphi)=(2\pi)^{-1}$ (fig. \ref{fig:Fig_Angle_Time_Distributions}A). 
Sampling independent and identically distributed random variables from these turning angle and speed distributions according to Eqs.~\ref{eq:overd1ds} and \ref{eq:overd2ds} by converting speeds into step lengths per time interval defines a time-discrete two-dimensional Ross-Pearson random walk in the comoving frame. Alternatively, one could sample the speeds from a (half-)Gaussian, as often done in the literature \citeMain{lenz2013constructing,Sant20} and in our experiments (Figs. \ref{fig:Fig2} and fig. \ref{fig:Fig_Angle_Time_Distributions}), which due to the central limit theorem also yields Gaussian position distributions in the long-time limit, like the Ross-Pearson random walk.

\subsection*{Constant speed dynamics} \label{si:csd}
In our experiments, a microrobot moves with a speed that is constant on average, and we can control the flight time $\tau_n$ during which it moves ballistically in the same direction (Fig. \ref{fig:Fig1}). Given some stochastic dynamics with variable speed $v_n$ defined by
Eqs.~\ref{eq:overd1ds} and \ref{eq:overd2ds}, our goal is to construct a corresponding stochastic process with constant speed that, for clarity of notation, here we denote as $v_{\rm c}={\rm const.}$, which exactly preserves the topology of the paths generated by the original variable speed process. That is, a microrobot is supposed to move with constant speed on exactly the same paths generated by exactly the same step-length distribution for both dynamics. This implies that, instead of using constant flight times $\tau_n=\tau_{\rm c}={\rm const.}$ for the variable speed dynamics, we use variable flight times $\tau_n$ for the constant speed dynamics by preserving the corresponding step lengths $\ell_n$ for both processes. Using the very same step lengths $\ell_n$ provides the crucial link between both dynamics. To do so, we need to transform Eq.~\ref{eq:overd2ds} such that we can sample independently and identically distributed random variables from a distribution of flight times $\xi_{\tau,n}\sim\rho(\tau)$ instead of a speed distribution $\xi_{v,n}\sim\rho(v)$ (Eq.~\ref{eq:overd2d2} in the main text). 

We can perform this transformation as follows. Consider a stochastic process for the speed $v_n=\xi_{v,n}$, where the random variable $\xi_{v,n}$  is drawn independently and identically distributed from a given speed distribution $\rho(v)$, $\xi_{v,n}\sim\rho(v)$. Then the distance $\ell_n$ traveled during a constant flight time $\tau_{\rm c}$ is given by
\begin{equation}
    \ell_n=v_n\tau_{\rm c}\: . \label{eq:vlteq}
\end{equation}
This defines $\ell_n=\xi_{\ell,n}$ as a new, transformed random variable for which we need to calculate the associated step length distribution $\xi_{\ell,n}\sim\rho(\ell)$. This can be done again by conservation of probability (cf.\ Eq.~\ref{eq:cop}), $\rho(v)dv=\rho(\ell)d\ell$,
where the Jacobian is calculated from Eq.~\ref{eq:vlteq} yielding
\begin{equation}
  \rho(\ell) =\frac{1}{\tau_{\rm c}}\rho(v)\: .  \label{eq:sld}
\end{equation}
This completes the transformation from $v_n=\xi_{v,n}\sim\rho(v)$ to the associated stochastic process $\ell_n=\xi_{\ell,n}\sim\rho(\ell)$ at constant flight times $\tau_{\rm c}$. 

Now assume that $v_n=v_{\rm c}= {\rm const.}$ instead of $\tau_n=\tau_{\rm c}={\rm const.}$, as in our experiments (Fig. \ref{fig:Fig1} and fig. \ref{fig:Fig_Instant_Speed}). We wish to preserve the step length $\ell_n$ at a given discrete time step $n$, as calculated above, so that the overall topology of a random path is preserved. That is, a microrobot traverses the very same distance $\ell_n$ for the very same turning angle $\varphi_n$ at a time step $n$ at now constant speed $v_{\rm c}$, paying the price that it will now do so at a typically different (longer or shorter) associated variable flight time $\tau_n$, making up for the corresponding original variable speed $v_n$. This is exactly the same idea underlying the transformation between L\'evy flights and L\'evy walks \citeMain{KRS08,ZDK15}, which we implement here within the comoving frame. Accordingly, we now remove the assumption that $\tau_n=\tau_{\rm c} = {\rm const.}$ in Eq.~\ref{eq:vlteq}, allowing again for variable flight times $\tau_n$, by implementing instead the constant speed constraint $v_n=v_{\rm c}={\rm const.}$ This leads to the complementary equation
\begin{equation}
    \tau_n=\frac{\ell_n}{v_{\rm c}} \: , \label{eq:vlteqc}
\end{equation}
for the very same $\ell_n$ as above, where $\tau_n=\xi_{\tau,n}$ is sampled from a flight time distribution $\xi_{\tau,n}\sim\rho(\tau)$. The latter we can obtain from the step length distribution $\rho(\ell)$, Eq.~\ref{eq:sld}, again by conservation of probability. Calculating the involved Jacobian by using Eq.~\ref{eq:vlteqc} gives
\begin{equation}
  \rho(\tau)=v_{\rm c}\rho(\ell)\label{eq:lvt}\: .
\end{equation}
Feeding Eq.~\ref{eq:sld} into Eq.~\ref{eq:lvt} yields the desired transformation between a given speed distribution at constant flight time $\tau_{\rm c}$ and the corresponding flight time distribution at constant speed $v_{\rm c}$,
\begin{equation}
    \rho(\tau)=\frac{v_{\rm c}}{\tau_{\rm c}}\rho(v)\: . \label{eq:ftpdf}
\end{equation}
The step lengths $\ell_n$ corresponding to flight times $\tau_n$ at constant speed $v_{\rm c}$ can be obtained from
Eq.~\ref{eq:vlteqc} yielding
\begin{equation}
    \ell_n=v_{\rm c}\xi_{\tau,n} \: ,
\end{equation}
which gives Eq.~\ref{eq:overd2d2} in the main text.

As an example of applying this transformation scheme from variable speed at constant flight times to constant speed at variable flight times, consider the Rayleigh distribution for the speed mentioned at the end of the previous section. We wish to transform this distribution into a corresponding flight time distribution by assuming now a constant speed $v_{\rm c}$. We can do so by solving Eq.~\ref{eq:vlteq} for $v_n$, solving Eq.~\ref{eq:vlteqc} for $\ell_n$, and replacing $\ell_n$ in the former equation by the latter yielding
\begin{equation}
     v_n=\frac{v_{\rm c}}{\tau_{\rm c}}\tau_n \: . \label{eq:vttrafo}
\end{equation}
Applying this change of variables to the stationary Rayleigh speed distribution Eq.~\ref{eq:rayleigh} yields for the associated stationary flight time distribution at constant speed
\begin{equation}
    \rho(\tau)=\frac{\tau}{\sigma_\tau^2} e^{-\frac{\tau^2}{2\sigma_\tau^2}}\: , \label{eq:ftpdfcs}
\end{equation}
with scale parameter $\sigma_\tau=\tau_{\rm c}\sigma_v/v_{\rm c}$, which is also a Rayleigh distribution. An $\alpha$-stable L\'evy speed distribution can be transformed into a corresponding $\alpha$-stable L\'evy flight time distribution exactly along the same lines. Since an exact definition for all parameter values $\alpha = 3 - \mu$ can only be given through its characteristic function in Fourier space, here we focus on the asymptotic representation in the form of power-law tails \citeMain{KRS08,ZDK15},
\begin{equation}
\rho(v) \approx C_1(\alpha)\;v^{-(\alpha + 1)} \:(v\to\infty) \label{eq:levypdf}
\end{equation}
with $C_1(\alpha)=(1/\pi)\sin(\pi\alpha/2)\Gamma(1+\alpha)$. Applying Eq.~\ref{eq:vttrafo} then gives
\begin{equation}
\rho(\tau) \approx C_1(\alpha)\frac{v_{\rm c}}{\tau_{\rm c}}\;\tau^{-(\alpha + 1)} \: (\tau\to\infty) \: ,
\end{equation}
yielding the functional form of Eq.~\ref{eq:powertrunc}, which was used for the experiments. 

We emphasize that, while this general transformation between variable and constant speed is designed to preserve the topology of a given random path by preserving the step length generated at any discrete time step, it strictly yields a pair of variable- and constant-speed stochastic processes that are {\em not equivalent} to each other, see Eqs.~\ref{eq:overd1d} and \ref{eq:overd2d} and Eqs.~\ref{eq:overd1d} and \ref{eq:overd2d2} in the main text. Clearly, for the former the speed distribution is not a $\delta$-function while for the latter, by definition, it is, and vice versa for the corresponding flight time distributions. This non-trivial connection and its consequences for the corresponding stochastic properties have been amply explored in the literature, e.g., for L\'evy flights and walks \citeMain{KRS08,ZDK15}.

Finally, choosing for $\xi_{\varphi,n}\sim\rho(\varphi)$ in Eq.~\ref{eq:overd1d} a uniform distribution and for $\xi_{\tau,n}\sim\rho(\tau)$ in Eq.~\ref{eq:overd2d2} a half-Gaussian distribution yields normal diffusion, while choosing for $\xi_{\varphi,n}\sim\rho(\varphi)$ in Eq.~\ref{eq:overd1d} a uniform distribution and for $\xi_{\tau,n}\sim\rho(\tau)$ in Eq.~\ref{eq:overd2d2} an $\alpha$-stable L\'evy distribution defines a uniform L\'evy walk in the comoving frame (Figs. \ref{fig:Fig1}, \ref{fig:Fig2} and \ref{fig:Fig3}) \citeMain{zaburdaev2016superdiffusive}.

\subsection*{Fractional Brownian motion and walks} \label{si:fbmcm}
The previous transformation between variable-speed and constant-speed dynamics applies to Markovian stochastic processes, i.e. without memory between different steps $\ell_n$. In our experiments, we also want to generate spatiotemporally correlated dynamics. A paradigmatic example of such a process is fractional Brownian motion (FBM). Therefore, in this section, we first discuss how to generate an analogue of two-dimensional FBM in a comoving frame under the constraint of constant speed to then explain our specific experimental implementation (Fig. \ref{fig:Fig4}) in the following section (\emph{Experimental implementation of fractional Brownian walks}). FBM is a Gaussian power-law correlated stochastic process that generates the whole spectrum of anomalous diffusion under parameter variation, from subdiffusion to
superdiffusion through normal diffusion \citeMain{mandelbrot1968fractional,KKK21}. While L\'evy walks yield superdiffusion due to sampling the step lengths $\ell_n$ from power-law distributions instead of Gaussians (Figs. \ref{fig:Fig1} and \ref{fig:Fig2}), FBM defines a fundamentally different class of anomalous stochastic process, where anomalous diffusion is due to a non-Markovian, power-law correlation decay of the velocity autocorrelation function (VACF) while sampling the velocities from Gaussian distributions \citeMain{BHOZ08}$,\,$\!\citeSI{embrechts2009selfsimilar}.

FBM is a continuous process in space and time, which can be defined in terms of overdamped Langevin dynamics driven by fractional Gaussian noise (FGN) \citeMain{jeon2010fractional,KKK21}. In
one dimension, the corresponding equation reads
\begin{equation}
\frac{dx}{dt}=\xi_{\rm FGN}(t)\: ,
\end{equation}
where $\xi_{\rm FGN}(t)$ holds for fractional, i.e. power-law correlated
Gaussian noise \citeMain{jeon2010fractional,KKK21}
\begin{equation}
\langle\xi_{\rm FGN}(t_1)\xi_{\rm FGN}(t_2)\rangle=2K_HH\Big[(2H-1)|t_1-t_2|^{2H-2}+2|t_1-t_2|^{2H-1}\delta(t_1-t_2)\Big] \label{eq:cffbm}
\end{equation}
with Hurst exponent $0<H<1$ and generalized diffusion coefficient $K_H$. The correlation function of the noise is thus identical with the VACF of the FBM process, which decays as a power law in time. From this equation, the mean squared displacement of FBM can be calculated
to \citeMain{jeon2010fractional,KKK21}
\begin{equation}
\langle x^2(t)\rangle=2K_Ht^{2H}\: .
\end{equation}
Under variation of the power-law exponent $\mu=2H$, FBM displays the whole spectrum of anomalous diffusion. 

For our experiments, we need to define a process at constant speed in the comoving frame that preserves the topology of the paths generated by FBM. Since, to our knowledge, no definition of such a stochastic process exists in this frame, we design a protocol to perform the required transformation. In line with our experimental set-up, we define this protocol in terms of time- and space-discrete FBM \citeMain{KKK21}$,\,$\!\citeSI{MoLi97}. The crucial aspect for this transformation is to capture the non-Markovian correlations in\ Eq.~\ref{eq:cffbm} inherent to FBM. To do so, we consider a whole time series of a time-discrete FBM path (e.g. as generated in simulations) by recording the exact sequence of speeds and associated turning angles $(v_n,\varphi_n)$. We then use our change of variables in Eq.~\ref{eq:vttrafo} to generate a Cartesian constant-speed version of FBM, defining the corresponding sequence of flight times and associated turning angles $(\tau_n,\varphi_n)$. In addition to preserving the step lengths at any time step $n$, reproducing the velocity correlations requires us to also preserve the turning angles between pairs of velocities at subsequent $n$. Using the resulting full sequence of flight times and turning angles $(\tau_n,\varphi_n)$ at constant speed indeed enables us both to maintain the initial correlation (as we show below) and to reproduce the topology of the original variable speed FBM in the comoving frame. We can consider this resulting constant-speed analogue of FBM as a \emph{fractional Brownian walk}, in analogy to transforming  L\'evy flights into L\'evy walks \citeMain{ZDK15}. The detailed protocol achieving this transformation from variable-speed to constant-speed FBM is defined as follows:

\begin{enumerate}
\item Generate two-dimensional time-discrete Cartesian FBM by a standard numerical algorithm.
\item Extract the sequence of speeds $v_n$ and turning angles $\varphi_n$ from a path of this numerically generated FBM at any time step $t_n=n\tau_{\rm c}$ for given $\tau_{\rm c} = {\rm const.}$ 
\item For a given $v_n$ in the sequence, calculate a corresponding flight time $\tau_n$ according to Eq.~\ref{eq:vttrafo} by setting the desired speed $v_{\rm c} = {\rm const.}$
\item Reproduce the sequence of $(\tau_n,\varphi_n)$ pairs in the comoving frame with constant speed $v_{\rm c}$, thus obtaining a constant-speed analogue of FBM in this frame that can be implemented with Eqs.~\ref{eq:overd1d} and \ref{eq:overd2d2}.
\end{enumerate}

The velocities of this process are no longer Gaussian-distributed nor are the speeds Rayleigh-distributed. However, according to Eq.~\ref{eq:vttrafo}, we can expect the flight-time distribution of this process at constant speed (corresponding to a variable-speed Gaussian stochastic process like FBM at constant flight times) to be Rayleigh-distributed (Eq. ~\ref{eq:ftpdfcs}), as confirmed experimentally (fig.~\ref{fig:Fig_Time_Distributions_fBm}).
Most importantly, we expect this FBM-like process to still preserve the non-Markovian VACF decay (Eq.~\ref{eq:cffbm}) as a defining feature of the original Gaussian FBM process. The latter property can be verified by starting from the definition of the VACF for an arbitrary non-constant speed, two-dimensional time-discrete process, given by
\begin{equation}
\langle \mathbf{v}_0\mathbf{v}_n\rangle=\langle v_0v_n\cos \varphi_n\rangle \: ,
\end{equation}
where, for $\mathbf{v}_0\parallel x$, the angle $\varphi_n$ is the polar angle between the velocity vector $\mathbf{v}_0$ at time step $n=0$ and the velocity vector $\mathbf{v}_n$ at time step $n$, with the angular brackets defining a suitable (time or ensemble) average. Assuming that the initial speed $v_0={\rm const.}$, we can rewrite
\begin{equation}
\langle \mathbf{v}_0\mathbf{v}_n\rangle=v_0\langle v_n\cos \varphi_n\rangle \: .
\end{equation}
As the polar coordinates, speed $v_n$ and orientation $\varphi_n$, for
the velocity vector $\mathbf{v}_n$ are decoupled,
\begin{equation}
\langle \mathbf{v}_0\mathbf{v}_n\rangle=v_0\langle
v_n\rangle\langle\cos \varphi_n\rangle \: .
\end{equation}
Because, by definition, our protocol preserves all angles $\varphi_n$ of the original time-discrete FBM process, we conclude that the VACF of the underlying original FBM process we started from is preserved if the constant speed $v_{\rm c}=v_0$ fulfills the equation 
\begin{equation}
v_{\rm c}=\langle v_n\rangle\: . \label{eq:cscon}
\end{equation}
For the speed-constant process, by averaging over Eq.~\ref{eq:vttrafo} and using Eq.~\ref{eq:cscon} we furthermore obtain
\begin{equation}
    \tau_{\rm c}=\langle\tau_n\rangle\: . \label{eq:cstau}
\end{equation}
This relationship provides an important matching of timescales, indicating that the decay of the original FBM VACF can only be observed at times longer than the cutoff time defined by $\tau_{\rm c}$. 

\subsection*{Experimental implementation of fractional Brownian walks} \label{si:scaling}
In our protocol above, we first need to generate FBM via a standard numerical algorithm, e.g. via the Davies-Harte method as done in our case \citeMain{davies1987tests}. However, the associated discretization of FBM \citeSI{MoLi97} generates spurious dependencies on the Hurst exponent $H$, which propagate into this method \citeMain{davies1987tests}. To avoid introducing such dependencies, we initially simulated space-time-discrete FBM via the Davies-Harte algorithm 
setting the constant flight time ($\tilde{\tau}_{\rm c}$) and the scale parameter ($\sigma_{\tilde{v}}$) to unitary values instead of directly setting the values needed for our experiments. Note that we now denote all variables generated in simulations with a tilde while all experimentally reproduced ones are still written without it.  To match our experimental length scales, we then rescaled the distance $\tilde{\ell}_n$ traveled in simulations during the constant flight time $\tilde{\tau}_{\rm c}$ by a factor $\kappa$, thus giving the actual experimental step length $\ell_n$ for our microrobots as
\begin{equation}
    \ell_n=\kappa\tilde{\ell}_n \label{eq:lscale} \, .
\end{equation}
We can therefore redefine all the relevant quantities discussed in the previous sections by taking this scaling into account. For brevity, in the following we just give the main results. After rescaling, Eq.~\ref{eq:vlteq} yields 
\begin{equation}
    v_n=\frac{\kappa\tilde{\ell}_n}{\tilde{\tau}_{\rm c}} \label{eq:vvscale}
\end{equation}
with scaled speed
\begin{equation}
v_n=\kappa \tilde{v}_n\:. \label{eq:vscale}
\end{equation}
Applied to Eq.~\ref{eq:rayleigh}, we obtain the scaled stationary Rayleigh speed distribution 
\begin{equation}
    \rho(v)=\frac{v}{\sigma_{v}^2}e^{-\frac{v^2}{2\sigma^2_{v}}} \label{eq:rscale}
\end{equation}
with scale parameter $\sigma_v=\kappa \sigma_{\tilde{v}}$. Next, the scaled Eq.~\ref{eq:vlteqc} reads
\begin{equation}
    \tau_n=\frac{\kappa\tilde{\ell}_n}{\tilde{v}_{\rm c}}\label{eq:tscale}
\end{equation}
with scaled flight time
\begin{equation}
    \tau_n=\kappa\tilde{\tau}_n\:.
\end{equation}
Combining Eqs.~\ref{eq:vvscale} and \ref{eq:tscale} leads to the scaled change of variables
\begin{equation}
     v_n=\frac{\tilde{v}_{\rm c}}{\tilde{\tau}_{\rm c}}\tau_n \: . \label{eq:vtscale}
\end{equation}
With this, we obtain the scaled Rayleigh flight time distribution
\begin{equation}
    \rho(\tau)=\frac{\tau}{\sigma_{\tau}^2} e^{-\frac{\tau^2}{2\sigma_{\tau}^2}}\: , \label{eq:ftscale}
\end{equation}
with scale parameter $\sigma_{\tau}=\tilde{\tau}_{\rm c}\sigma_{v}/\tilde{v}_{\rm c}=\kappa\sigma_{\tilde{\tau}}$. Note that this distribution is defined with respect to the original parameters $\tilde{\tau}_{\rm c}$ and $\tilde{v}_{\rm c}$ used in the simulations before scaling. Along the same lines, the scaled VACF is now given by
\begin{equation}
\langle \mathbf{v}_0\mathbf{v}_n\rangle=\kappa^2\langle\tilde{\mathbf{v}}_0\tilde{\mathbf{v}}_n\rangle \: .
\end{equation}
Following the same reasoning as before, we choose
\begin{equation}
    v_{\rm c}=\kappa \tilde{v}_{\rm c}=\kappa \tilde{v}_0 \label{eq:vctilde}
\end{equation}
yielding, as a condition for the scaled constant-speed FBM dynamics to preserve the topology of the original FBM process,
\begin{equation}
v_{\rm c}=\kappa\langle \tilde{v}_n\rangle=\langle v_n\rangle \label{eq:kappav}
\end{equation}
as the scaled analogue to Eq.~\ref{eq:cscon}. Likewise, for the relevant timescale of the scaled constant-speed process, we obtain 
\begin{equation}
    \tau_{\rm c}=\kappa\tilde{\tau}_{\rm c}=\kappa\langle\tilde{\tau}_n\rangle=\langle\tau_n\rangle \label{eq:kappatau}
\end{equation}
as the scaled analogue to Eq.~\ref{eq:cstau}. As before, Eq.~\ref{eq:kappatau} provides an important matching of timescales, indicating that the decay of the original FBM VACF can only be observed at times longer than the scaled cutoff time defined by $\kappa \tilde{\tau}_{\rm c}$. In our experiments, for sampling times $\delta t$ below this cut-off time, one can indeed observe the typical exponentially decaying short-time correlation due to the microrobots' short-term drift in the magnetic driving field \citeMain{Lowen2020}.

\clearpage
\begin{figure}[h!]
\centering
\includegraphics[width = 0.68\textwidth]{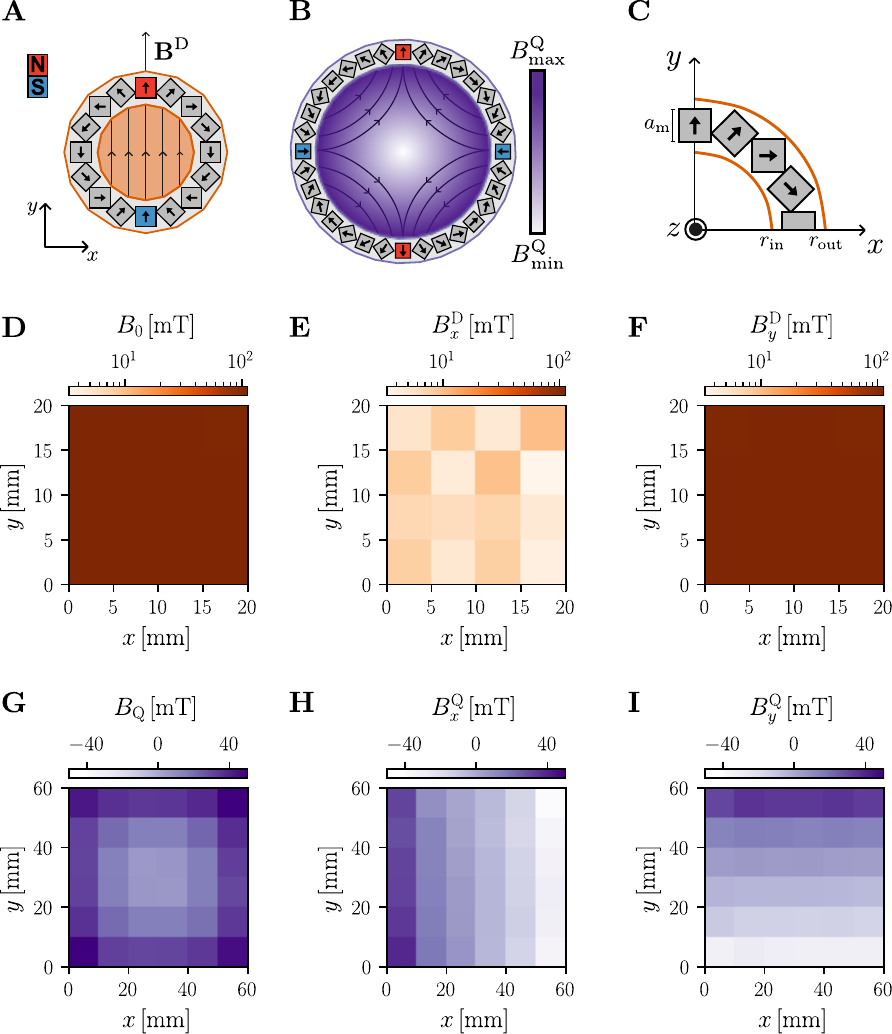}
\caption{{\bf Magnetic fields of the Halbach cylinders}. 
(\textbf{A}-\textbf{B}) In-scale schematics of a cylindrical Halbach (\textbf{A}) dipole and (\textbf{B}) quadrupole, built with cubic permanent magnets (gray squares with black arrows pointing to their north poles). The cylinders' magnetic north (N, red) and south (S, blue) poles as well as flux lines are shown. The strong dipolar field $\mathbf{B}^{\rm D}=B_0 \mathbf{e}_y$ is homogeneous and directed along the $y$-axis (unit vector, $\mathbf{e}_y$). The weaker quadrupolar field $\mathbf{B}^{\rm Q}$ consist of two orthogonal linear components. 
(\textbf{C}) The cubic magnets (side length, $a_{\rm m}$) are arranged in a circular pattern in the $xy$ plane,  with the cylinders' axis along $z$ (Experimental Section, table \ref{tab_magnets}).  
The inner radius $r_\mathrm{in}$ (from the center to the magnets’ inner edge) defines the experimental area. The outer radius $r_\mathrm{out}$ (from center to the magnets' outer edge) delimits the size of the cylinder.
(\textbf{D}) Measured dipolar magnetic field intensity $B_0$ with (\textbf{E}) $x$- and (\textbf{F}) $y$-components, $B^\mathrm{D}_x$ and $B^\mathrm{D}_y$. $B^\mathrm{D}_x$ is one order of magnitude smaller than $B^\mathrm{D}_y$. 
(\textbf{G}) Measured quadrupolar magnetic field intensity $B_{\rm Q}$ with (\textbf{H}) $x$- and (\textbf{I}) $y$-components, $B^\mathrm{Q}_x$ and $B^\mathrm{Q}_y$, which are orthogonal and linear in space.
}\label{fig:Fig_Halbach}
\end{figure}

\begin{figure}[h!]
\centering
\includegraphics[width = \textwidth]{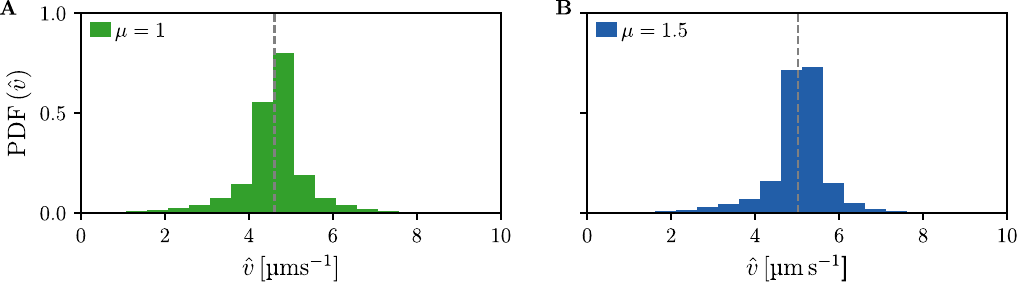}
\caption{{\bf Instantaneous microrobots' speed}. (\textbf{A}-\textbf{B}) Probability density functions (PDF) of the instantaneous microrobots' speed $\hat{v}$ extracted from the two trajectories in Fig. \ref{fig:Fig1}E for the cases of (\textbf{A}) normal diffusion ($\mu$ = 1) and (\textbf{B}) superdiffusion ($\mu$ = 1.5). The speed was calculated using a time window of $0.42 \,  {\rm s}$ (corresponding to five video frames) moving along the trajectory to minimize the impact of the tracking localization noise. The average speed (dashed vertical lines) is (\textbf{A}) $\langle\hat{v}\rangle = 4.61 \pm 0.82$~\um $\rm s^{-1}$ and (\textbf{B}) $\langle\hat{v}\rangle = 5.03\pm 0.74$~\um $\rm s^{-1}$.   
Due to the narrow distributions, we consider the microrobots to move approximately at constant speed in our experiments.}\label{fig:Fig_Instant_Speed}
\end{figure}

\begin{figure}[h!]
\centering
\includegraphics[width = \textwidth]{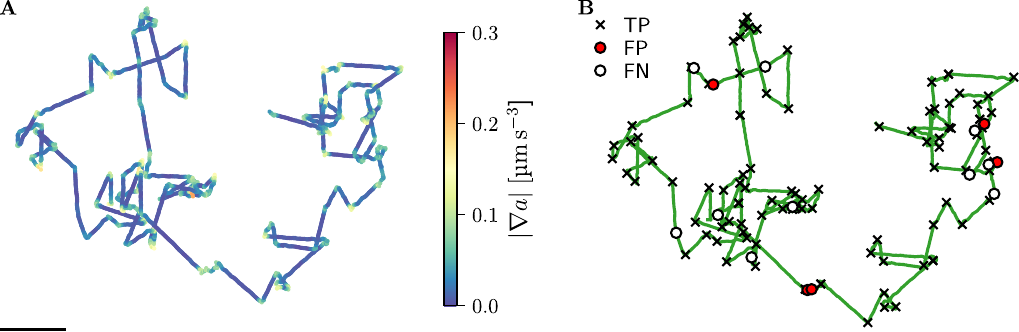}
\caption{{\bf Detection of turning points on microrobots' trajectories}. (\textbf{A}) Portion of the normal diffusion trajectory in Fig. \ref{fig:Fig1}E with the absolute values of the acceleration magnitude gradient ($|\nabla |\mathbf{a}||$) superimposed. Peaks in this quantity identify the turning points (Experimental Section).
(\textbf{B}) The detected turning points are validated against the quadrupole rotations by calculating the F1 score from true positives (crosses, TP), false positives (filled circles, FP), and false negatives (empty circles, FN) (Experimental Section). Scale bar: 25~\um.} \label{fig:Fig_Turning_Points}
\end{figure}

\begin{figure}[h!]
\centering
\includegraphics[width = \textwidth]{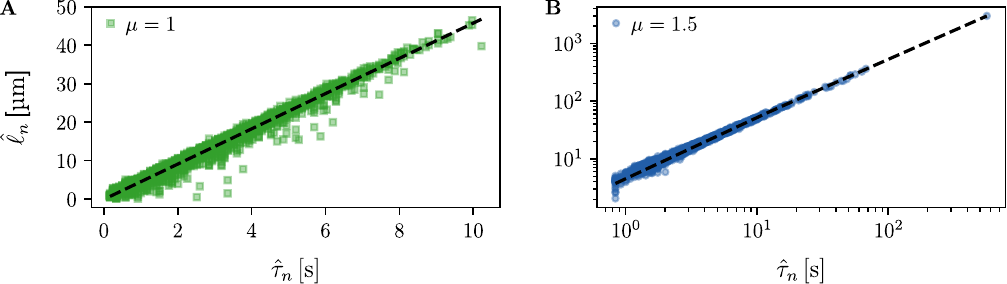}
\caption{{\bf Linearity between step lengths and flight times}. Microrobots' step lengths $\hat{\ell}_n$ as a function of flight times $\hat{\tau}_n$ between turns extracted from the two trajectories in Fig. \ref{fig:Fig1}E for the cases of (\textbf{A}) normal diffusion ($\mu$ = 1) and (\textbf{B}) superdiffusion ($\mu$ = 1.5). The linear fits to the data (black dashed lines) confirms that the microrobots move at an approximately constant speed for all step lengths. The slope of the fit lines provides an estimate of the average speed of the microrobots alternative to the instantaneous velocity (fig. \ref{fig:Fig_Instant_Speed}):  (\textbf{A}) $\langle\hat{v}\rangle = 4.585 \pm 0.009$ \um~\, $\rm s^{-1}$ and (\textbf{B}) $\langle\hat{v}\rangle = 5.298 \pm 0.002$ \um~$\rm s^{-1}$. These values are in good agreement with those in fig. \ref{fig:Fig_Instant_Speed}.}\label{fig:Fig_Time_Step_Speed}
\end{figure}

\begin{figure}[h!]
\centering
\includegraphics[width = \textwidth]{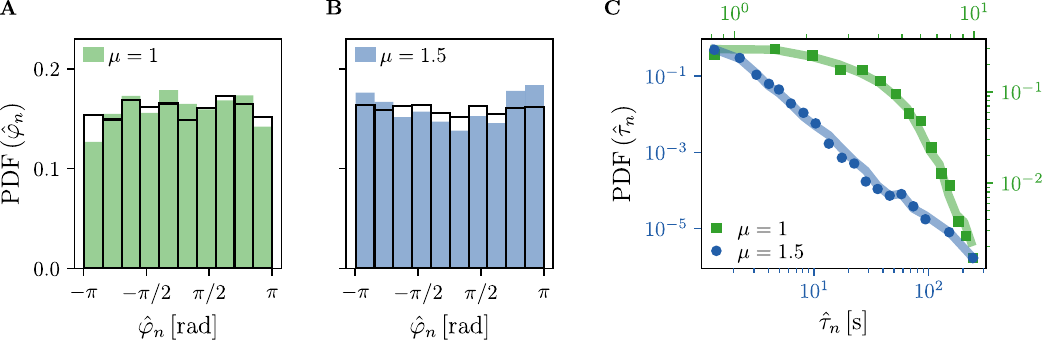}
\caption{{\bf Distributions of microrobots' turning angles and flight times}. 
\textbf{(A-B)} Probability density functions (PDF) of the turning angles {$\hat{\varphi}_n$} extracted from the two trajectories in Fig. \ref{fig:Fig1}E for \textbf{(A)} normal diffusion ($\mu$ = 1) and \textbf{(B)} superdiffusion ($\mu$ = 1.5). The PDFs of twice the quadrupole rotation angles ($\varphi_n = 2\Delta\beta$) are represented with black histograms in (\textbf{A}-\textbf{B}), indicating  uniform distributions on the circle and that ${\rm PDF} (\hat{\varphi}_n) \sim {\rm PDF} ({\varphi}_n) = {\rm PDF} (2\Delta\beta)$ (Experimental Section).
\textbf{(C)} PDFs of microrobot's flight times $\hat{\tau}_n$ (as in Fig. \ref{fig:Fig2}A) for the two trajectories in Fig. \ref{fig:Fig1}E corresponding to normal diffusion ($\mu$ = 1, green squares) and superdiffusion ($\mu$ = 1.5, blue circles). The thick background lines are the PDFs of the quadrupole rotation times $\tau_n$, showing that ${\rm PDF} (\hat\tau_n) \sim {\rm PDF} (\tau_n)$. The axis colors reflect those of the respective distributions.}
\label{fig:Fig_Angle_Time_Distributions}
\end{figure}

\begin{figure}[h!]
\centering
\includegraphics[width = \textwidth]{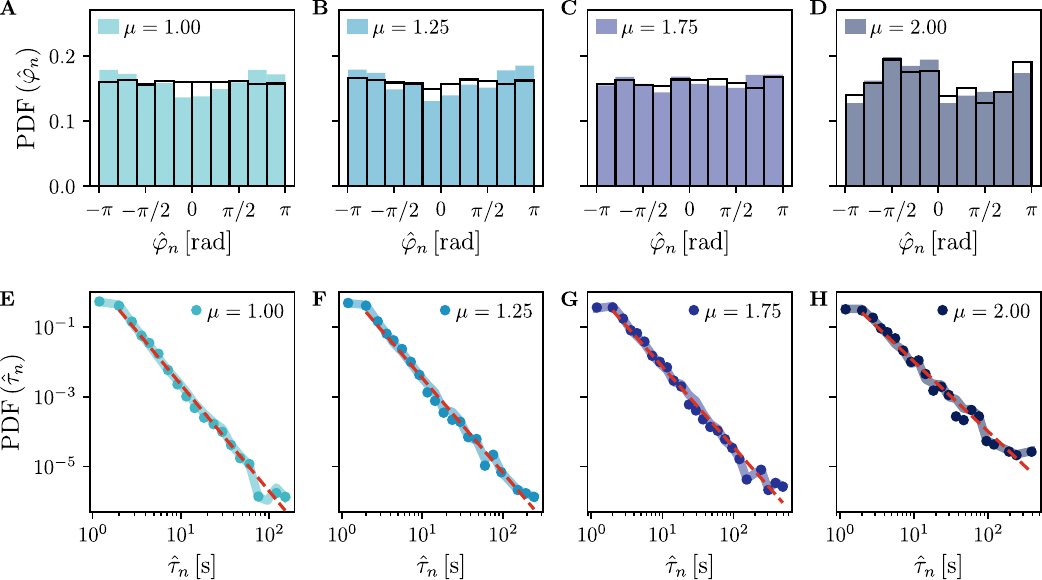}
\caption{{\bf Distributions of microrobots' turning angles and flight times for L\'evy walks.} Probability density functions (PDF) of (\textbf{A}-\textbf{D}) microrobots' turning angles $\hat{\varphi}_n$ and {(\textbf{E}-\textbf{H})} flight times $\hat{\tau}_n$ for the trajectories in Fig. \ref{fig:Fig3}A yielding L\'evy walks with different anomalous diffusion exponents $\mu$. 
In (\textbf{A}-\textbf{D}), the PDFs of twice the quadrupole rotation angles ($\varphi_n = 2\Delta \beta$) are represented with black histograms, indicating uniform distributions on the circle and that ${\rm PDF} (\hat{\varphi}_n) \sim {\rm PDF} ({\varphi}_n) = {\rm PDF} (2\Delta\beta)$ (Experimental Section).
In (\textbf{E}-\textbf{H}), the data are fitted to power laws (red dashed lines) showing consistent scaling ($\sim \hat{\tau}_n^{\hat{\mu}-4}$) with L\'evy walk models (table \ref{tab_Stats_Fig3}). The thick background lines are the PDFs of the quadrupole rotation time $\tau_n$, showing that ${\rm PDF} (\hat\tau_n) \sim {\rm PDF} (\tau_n)$.}\label{fig:Fig_Angle_Time_Distributions_Levy}
\end{figure}

\begin{figure}[ht!]
\centering
\includegraphics[width = 0.5\textwidth]{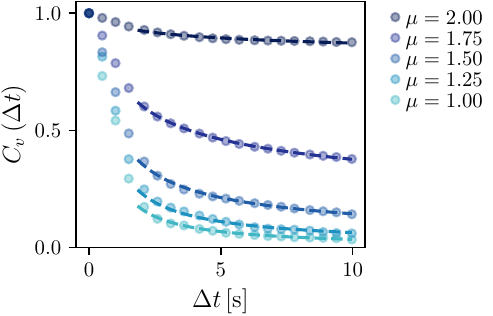}
\caption{{\bf Velocity autocorrelation function for L\'evy walks}. Normalized time-averaged velocity autocorrelation $C_v$ as a function of lag time $\Delta t$ for the trajectories in Fig. \ref{fig:Fig3}A, yielding L\'evy walks with different anomalous diffusion exponents $\mu$. Fitting the tail of the data with a power law (dashed lines) confirms the scaling  characteristic of L\`evy walks ($\sim \Delta t^{\hat{\mu}-2}$, table \ref{tab_Stats_Fig3}). 
}\label{fig:Cv_Levy}
\end{figure}

\begin{figure}[h!]
\centering
\includegraphics[width = \textwidth]{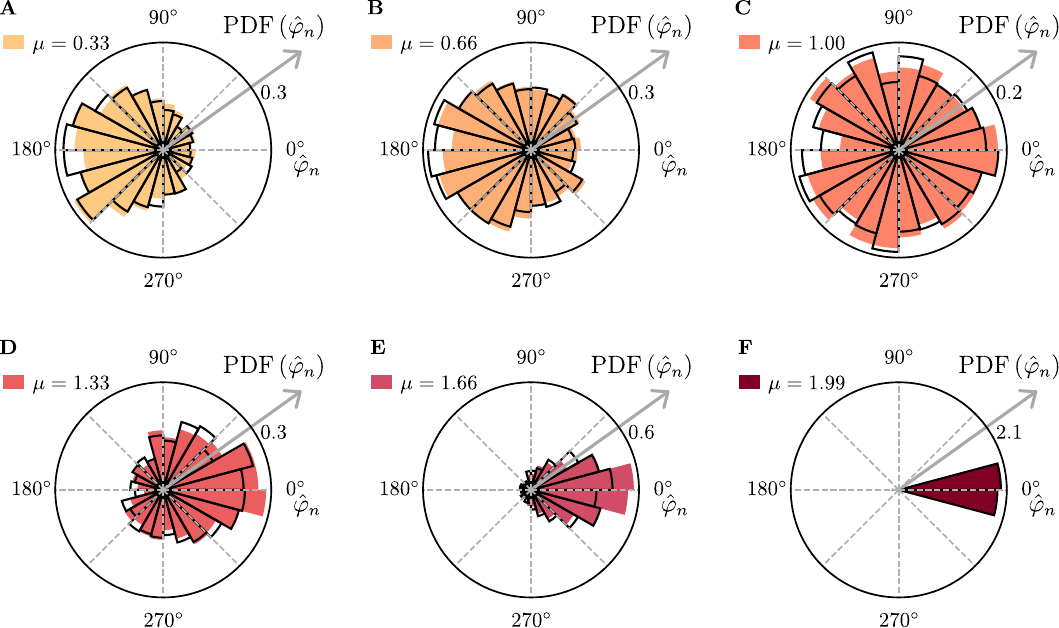}
\caption{{\bf Distributions of turning angles for fractional Brownian walks}. Polar histograms of the microrobots' turning angles $\hat{\varphi}_n$ from the trajectories in Fig. \ref{fig:Fig4}A showing the shift from backward (negative persistence) to forward propagation (positive persistence) as the anomalous diffusion exponent $\mu$ increases from subdiffusion (\textbf{A}: $\mu = 0.33$; \textbf{B}: $\mu = 0.66$) to diffusion (\textbf{C}: $\mu = 1$) to superdiffusion (\textbf{D}: $\mu = 1.33$; \textbf{E}: $\mu = 1.66$; \textbf{F}: $\mu = 1.99$).
(\textbf{A}-\textbf{B}) In subdiffusive motion ($\mu < 1$), microrobots tend to move in a direction opposite to their previous one (towards $180^{\circ}$). This tendency is stronger the lower the values of $\mu$. (\textbf{C}) For normal diffusion ($\mu = 1$), the uniform distribution of turning angles highlights the absence of directional correlations. (\textbf{D}-\textbf{F}) In superdiffusive regimes ($\mu > 1$), the microrobots are likely to maintain their current direction (towards $0^{\circ}$), as in directed motion. This tendency is stronger the higher the values of $\mu$. The black solid lines represent the distributions of twice the quadrupole rotation angle for reference.}
\label{fig:Polar_plots_fbm}
\end{figure}

\begin{figure}[h!]
\centering
\includegraphics[width = 0.65\textwidth]{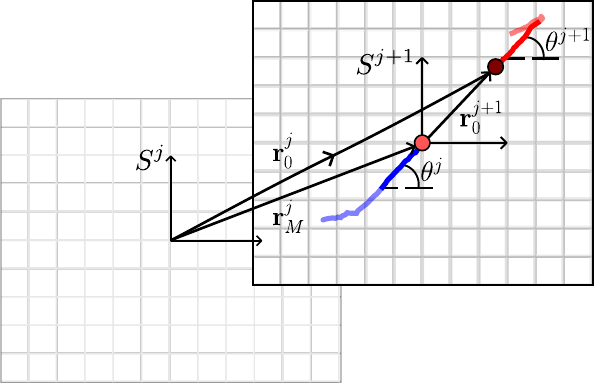}
\caption{{\bf Trajectory stitching}. Vectorial relationship (Eq. \ref{eq:stiching}) for the translation of the trajectory points $\mathbf{r}^{j+1}_i$ (red) acquired in the coordinate systems $S^{j+1}$ of the $(j+1)^{\rm th}$ video back to points $\mathbf{r}^{j}_i$ (blue) in the coordinate system $S^j$ of the $j^{\rm th}$ video to reconnect them to the previous part of the full microrobot's  trajectory. $\mathbf{r}_M^{j}$ is the position of the last recorded point in the $j^{\rm th}$ video in its reference system, and $\mathbf{r}^{j+1}_0$ and $\mathbf{r}^{j}_0$ respectively identify the first microrobot's position in the $(j+1)^{\rm th}$ video in its reference system and in the reference system of the $j^{\rm th}$ video. To facilitate interpolation of the stitched trajectory, any two consecutive videos respectively finished (dark blue line) and started (dark red line) with an at least 1-s long portion of the same step length in the full trajectory, with polar angles $\theta^j$ and $\theta^{j+1}$ in each respective reference system. The side length of each grid square corresponds to $5$~\um. 
}\label{fig:Fig_Traj_StitchingA}
\end{figure}

\begin{figure}[h!]
\centering
\includegraphics[width = 0.8\textwidth]{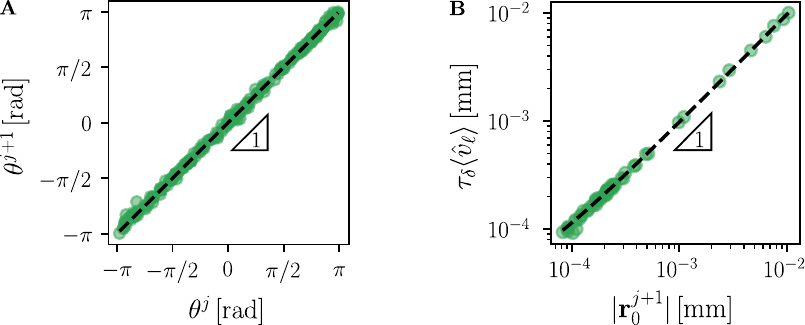}
\caption{{\bf Verification of trajectory stitching}.
(\textbf{A}) 
Linear relationship between $\theta^j$ and $\theta^{j+1}$ (as defined in fig. \ref{fig:Fig_Traj_StitchingA}) showing that $\theta^{j+1} \approx \theta^j$. A linear fit to $f(\theta^j) = k' \theta^j$ (dashed line) with slope $k' = 1.002 \pm 0.002$  confirms that the microrobot's direction of motion is preserved during the reconstruction of the trajectory.
(\textbf{B}) Linear relationship between direct measurements of the distance $|\mathbf{r}^{j+1}_0|$ (as defined in fig. \ref{fig:Fig_Traj_StitchingA}) from image analysis and measurements of stage displacement, and its estimate as $\tau_{\delta} \langle \hat{v}_\ell \rangle$, where $\langle \hat{v}_\ell \rangle$ is the mean microrobot's speed in the two reconnected portions from the same step length and $\tau_{\delta}$ is the time elapsed between recordings of two consecutive videos. A linear fit to $f(|\mathbf{r}^{j+1}_0|) = k' |\mathbf{r}^{j+1}_0|$ with slope $k' = 0.965 \pm 0.003$ confirms the good match between these two values. Linear trend with slope 1 shown for reference.}\label{fig:Fig_Traj_StitchingB}
\end{figure}

\begin{figure}[h!]
\centering
\includegraphics[width = 0.5\textwidth]{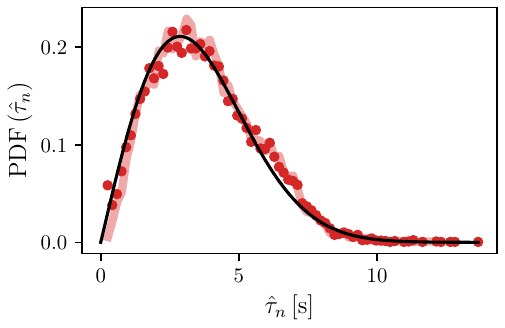}
\caption{{\bf Flight time distribution for fractional Brownian walks}. Probability density function (PDF) of microrobots' flight times ($\hat{\tau}_n$) for all fractional Brownian walks (Fig \ref{fig:Fig4}A). Data collapse on a Rayleigh distribution given by Eq. \ref{eq:ftscale} (black solid line), independent of the value of the anomalous diffusion exponent $\mu$. The thick background line is the PDF of the quadrupole rotation time $\tau_n$, showing that ${\rm PDF} (\hat\tau_n) \sim {\rm PDF} (\tau_n)$.}\label{fig:Fig_Time_Distributions_fBm} 
\end{figure}

\clearpage

\begin{table}[h]
\centering
\caption{{\bf Parameters of the Halbach cylinders.} Remanence $B_\mathrm{R}$ and relative permeability $\mu_\mathrm{R}$ of $k$ cubic neodymium magnets with side length $a_\mathrm{m}$ used to build two identical circular arrays for the Halbach dipole and one for the Halbach quadrupole for the experiments. The Halbach cylinders are characterized by the following parameters: inner radius $r_\mathrm{in}$, outer radius $r_\mathrm{out}$, average radius $r_\mathrm{c}$ and height $h = a_{\rm m}$.}\label{tab_magnets}
\begin{tabular}{lcccccccc}
\\
\hline
Halbach cylinder & $k$ & $B_\mathrm{R}$ [T] & $\mu_\mathrm{R}$ & $a_\mathrm{m}$ [mm] & $r_\mathrm{in}$ [mm] & $r_\mathrm{out}$ [mm] & $r_\mathrm{c}$ [mm] & $h$ [mm] \\
\hline
Dipole & 16 & 1.48 & 1.05 & 9.5 & 23.3 & 36.8 & 30.05 & 9.5\\
Quadrupole & 32 & 1.32 & 1.05 & 7 & 41.8 & 51.6 & 46.7 & 7\\
\hline
\end{tabular}
\end{table}

\begin{table}[h]
\centering
\caption{Estimated anomalous diffusion exponent $\hat{\mu}$ for L\'evy walks (Figs. \ref{fig:Fig2} and \ref{fig:Fig3}) based on fitting the mean squared displacement ($\hat{\mu}_{\rm MSD}$, Figs. \ref{fig:Fig1} and \ref{fig:Fig3}), the quadrupole time probability distribution function ($\hat{\mu}_{\tau_n}$, Fig. \ref{fig:Fig2}, figs. \ref{fig:Fig_Angle_Time_Distributions} and \ref{fig:Fig_Angle_Time_Distributions_Levy}), the flight time probability distribution function  ($\hat{\mu}_{\hat{\tau}_n}$, Fig. \ref{fig:Fig2}, figs. \ref{fig:Fig_Angle_Time_Distributions} and \ref{fig:Fig_Angle_Time_Distributions_Levy}), the step length probability distribution function  ($\hat{\mu}_{\hat{\ell}_n}$, Figs. \ref{fig:Fig2} and \ref{fig:Fig3}), and the velocity autocorrelation function ($\hat{\mu}_{C_v}$, fig. \ref{fig:Cv_Levy}). The first column shows the corresponding ground truth value of $\mu$. Corresponding instantaneous microrobot's speed $\langle\hat{v}\rangle$ are also shown. 
Errors represent one standard deviation of the respective fit parameters.}\label{tab_Stats_Fig3}

\begin{tabular}{ccccccc}
\\
\hline
$\mu$ & $\langle\hat{v}\rangle$~[\um~$\rm s^{-1}]$ & $\hat{\mu}_{\rm MSD}$ & $\hat{\mu}_{\tau_n}$ & $\hat{\mu}_{\hat{\tau}_n}$ & $\hat{\mu}_{\hat{\ell}_n}$ & $\hat{\mu}_{C_v}$ \\
\hline
$1.00$ & $5.03\pm0.97$ & $1.0509\pm0.0002$ & $0.93\pm0.09$ & $0.94\pm0.08$ & $0.95\pm0.06$ & $1.039\pm0.010$ \\
$1.25$ & $4.88\pm0.95$ &$1.2578\pm0.0003$ & $1.30\pm0.05$ & $1.29\pm0.07$ & $1.22\pm0.07$ & $1.200\pm0.013$ \\
$1.50$ & $5.03\pm0.74$ &$1.5473\pm0.0004$ & $1.49\pm 0.04$  & $1.46\pm0.06$ & $1.45\pm0.05$ & $1.434\pm0.005$ \\
$1.75$ & $5.11\pm0.75$ &$1.7498\pm0.0003$ & $1.71\pm0.06$ & $1.70\pm0.06$ & $1.71\pm0.09$ & $1.704\pm0.001$ \\
$2.00$ & $4.46\pm0.93$ &$1.9158\pm0.0003$ & $2.05\pm0.06$ & $2.00\pm0.08$ & $2.03\pm0.07$ & $1.964\pm0.001$ \\
\hline
\end{tabular}
\end{table}

\begin{table}[h]
\centering
\caption{Estimated anomalous diffusion exponent $\hat{\mu}$ for fractional Brownian walks (Fig. \ref{fig:Fig4}) based on fitting the mean squared displacement ($\hat{\mu}_{\rm MSD}$) and the velocity autocorrelation function ($\hat{\mu}_{C_v}$). The first column shows the corresponding ground truth value of $\mu$. Corresponding instantaneous microrobot's speed $\langle\hat{v}\rangle$ are also shown. Errors represent one standard deviation of the respective fit parameters.}\label{tab_Stats_Fig4}
\begin{tabular}{cccc}
\\
\hline
$\mu$ & $\langle\hat{v}\rangle$~[\um~$\rm s^{-1}]$ & $\hat{\mu}_{\rm MSD}$ & $\hat{\mu}_{C_v}$ \\
\hline
$0.33$ & $3.92\pm0.76$ & $0.36944\pm0.00030$ & $0.29\pm0.08$ \\
$0.66$ & $4.99\pm0.84$ & $0.67492\pm0.00028$ & $0.67\pm0.09$ \\
$1.00$ & $5.66\pm0.82$ & $0.95485\pm0.00022$ & $1.00\pm0.08$ \\
$1.33$ & $4.98\pm0.78$ & $1.29699\pm0.00023$ & $1.27\pm0.06$ \\
$1.66$ & $4.60\pm0.91$ & $1.67541\pm0.00021$ & $1.60\pm0.05$ \\
$1.99$ & $3.71\pm0.56$ & $1.98648\pm0.00003$ & $2.00\pm0.06$ \\
\hline
\end{tabular}
\end{table}

\begin{raggedright}

\end{raggedright}

\end{document}